\documentclass[12pt]{article}
\usepackage{graphicx}
\usepackage{float}
\usepackage{makecell} %multiple line in a table cell

\usepackage{soul} %highlight text
\usepackage{pbox} %multiple line in a table cell
\usepackage{subcaption}
\usepackage[dvipsnames]{xcolor}
\usepackage{stmaryrd}
\usepackage{todonotes}
\usepackage[skip=10pt plus1pt, indent=20pt]{parskip}
\usepackage{amsmath}
\usepackage{amssymb}
\usepackage{indentfirst} %indent after a section
\usepackage{geometry}
\usepackage[utf8]{inputenc}
\usepackage{lscape} %landscape table
\usepackage{float} %figure position fixed
\usepackage{tensor}
 \usepackage[percent]{overpic}  %reference over figure
\usepackage[T1]{fontenc}
\usepackage[export]{adjustbox}
\usepackage{tikz-cd}
\usepackage[outline]{contour}
\usepackage{lipsum,tikz}
\usepackage{hyperref}

\usepackage[backend=bibtex,style=numeric-comp,natbib=true]{biblatex} % Use the bibtex backend with the 'numeric'/'authoryear' citation style (which resembles APA) (Template)
\usepackage{tabularx,ragged2e,booktabs}
\usepackage{authblk}
\usepackage{tcolorbox} % create box
\usepackage{algorithm} %algorithm environment
\usepackage[noend]{algpseudocode}

\newcommand{\pv}{\textit{P. vivax}}
\newcommand{\pf}{\textit{P. falciparum}}
\newcommand{\etal}{\textit{et al. }}
\begin{document}

\title{Mathematical models of \textit{Plasmodium vivax} transmission: a scoping review}
%\title{A scoping review of mathematical models of \textit{Plasmodium vivax}}

\author[1,2]{Md Nurul Anwar}
\author[3,4]{Lauren Smith}
\author[5,6]{Angela Devine}
\author[1]{Somya Mehra}
\author[1]{Camelia R. Walker}
\author[1]{Elizabeth Ivory}
\author[3,4]{Eamon Conway}
\author[3,4]{Ivo Mueller}
\author[1,7]{James M. McCaw}
\author[1]{Jennifer A. Flegg}
\author[1,8,9]{Roslyn I. Hickson}

\affil[1]{School of Mathematics and Statistics, The University of Melbourne, Parkville, Australia}
\affil[2]{Department of Mathematics, Bangabandhu Sheikh Mujibur Rahman Science and Technology University, Gopalganj 8100, Bangladesh}
\affil[3]{The Walter and Eliza Hall Institute of Medical Research, Melbourne, Victoria, Australia}
\affil[4]{Department of Medical Biology, University of Melbourne, Melbourne, Victoria, Australia}
\affil[5]{Division of Global and Tropical Health, Menzies School of Health Research, Charles Darwin University, Darwin, Australia}
\affil[6]{Health Economics Unit, Centre for Health Policy, Melbourne School of Population and Global
Health, The University of Melbourne, Parkville, Australia}
\affil[7]{Centre for Epidemiology and Biostatistics, Melbourne School of Population and Global
Health, The University of Melbourne, Parkville, Australia}

\affil[8]{Australian Institute of Tropical Health and Medicine, James Cook University, Townsville, Australia}
\affil[8]{Commonwealth Scientific and Industrial Research Organisation (CSIRO), Townsville, Australia}

\date{}                     %% if you don't need date to appear
\setcounter{Maxaffil}{0}
\renewcommand\Affilfont{\itshape\small}
\maketitle
\begin{abstract}
\textit{Plasmodium vivax} is one of the most geographically widespread malaria parasites in the world, primarily found across South-East Asia, Latin America, and parts of Africa. \textit{P. vivax} is unique compared to most other \textit{Plasmodium} parasites due to its ability to remain dormant in the human liver as hypnozoites and subsequently reactivate after the initial infection (i.e. relapse infections). Mathematical modelling approaches have been widely applied to understand \pv~dynamics and predict the impact of intervention outcomes. Models that capture \pv~dynamics differ from those that capture \pf~dynamics, as they must account for relapses caused by the activation of hypnozoites. In this article, we provide a scoping review of mathematical models that capture \pv~transmission dynamics published between January 1988 and May 2023. The primary objective of this work is to provide a comprehensive summary of the mathematical models and techniques used to model \pv~dynamics. In doing so, we aim to assist researchers working on mathematical epidemiology, disease transmission, and other aspects of \pv~malaria by highlighting best practices in currently published models and highlighting where further model development is required. We categorise \pv~models according to whether a deterministic or agent-based approach was used. We provide an overview of the different strategies used to incorporate the parasite's biology, use of multiple scales (within-host and population-level), superinfection, immunity, and treatment interventions. In most of the published literature, the rationale for different modelling approaches was driven by the research question at hand. Some models focus on the parasites' complicated biology, while others incorporate simplified assumptions to avoid model complexity.  Overall, the existing literature on mathematical models for \pv~encompasses various aspects of the parasite's dynamics. We recommend that future research should focus on refining how key aspects of \pv~dynamics are modelled, including spatial heterogeneity in exposure risk, the accumulation of hypnozoite variation, the interaction between \pf~and \pv, acquisition of immunity, and recovery under superinfection.  

\end{abstract}
% \begin{multicols}{2}

\section{Introduction}\label{Intro}
Malaria remains a  significant public health problem, with an estimated 247 million cases and 619,000 deaths reported worldwide in 2021 alone \cite{world2022world}. Malaria is most prevalent in the World Health Organisation (WHO) African Region, while the South-East Asia Region has the second-highest estimated malaria burden globally. \textit{Plasmodium vivax} is currently the most geographically widespread of the malaria parasites, resulting in significant associated global morbidity and mortality \cite{antinori2012biology,battle2019mapping,anwar2023optimal,price2020plasmodium}. \pv~has been responsible for approximately 45\% of malaria cases in the WHO South-East Asia Region since 2000 and is widely prevalent in countries across Asia, Latin America, and the Pacific Islands \cite{battle2019mapping,world2022world,price2020plasmodium}. \pv~has often been overlooked and mistakenly considered as ``benign'' in the past \cite{price2007vivax,price2020plasmodium}. More recent research has produced evidence that, in addition to causing severe illness, \pv~infection can cause long-term health consequences such as anaemia, impaired cognitive development, and chronic kidney disease \cite{breman2007defining,tjitra2008multidrug,kochar2009severe,baird2013evidence}. The economic impact of \pv~malaria is also significant, as the disease can lead to decreased productivity, increased healthcare costs, and reduced economic growth in endemic areas \cite{devine2021global}.  

Mathematical modelling is an important tool that allows us to understand dynamic systems in various fields ranging from physics and engineering to social sciences and biology \cite{keeling2011modeling}. Mathematical modelling can provide valuable insight into infectious disease dynamics and plays an important role in informing public health policy and decision-making \cite{barnabas2006epidemiology,huppert2013mathematical}. Infectious disease modelling has been widely used to understand the transmission of malaria, particularly \textit{Plasmodium falciparum}, and the impact of interventions to control and eliminate malaria \cite{mandal2011mathematical,smith2018agent}. Modelling of \pv~transmission differs from \pf~modelling, due to the need to account for recurrent infections caused by the activation of hypnozoites, a dormant liver stage of the parasite.

\pv~parasites are introduced into the human body through infectious \textit{Anopheles} mosquito bites. \pv~parasites then travel to the liver, where they undergo a series of developmental and replication stages \cite{ jones2006malaria,kebaier2009kinetics} before the liver-stage parasites are released into the blood, causing blood-stage infections. Individuals experiencing a blood-stage infection may become symptomatic, with symptoms such as fever and fatigue, or be asymptomatic. One of the significant characteristics of \pv~infection is that, as part of the parasites' life-cycle, they can remain dormant in the liver for weeks or months \cite{imwong2007relapses} as hypnozoites that can cause further blood-stage infections (called relapses) upon reactivation. Importantly, between 79 and 96\% of \pv~cases are due to relapses \cite{robinson2015strategies,huber2021radical,adekunle2015modeling,commons2020estimating}. It can be challenging to distinguish a relapse from other types of recurrent malaria, such as a reinfection (i.e. malaria due to a new infectious bite) or a recrudescence (i.e. recurrence of malaria due to incomplete elimination of blood-stage infections, often associated with treatment failure) \cite{ghosh2020mathematical}. Relapse dynamics typically follow temperate or tropical phenotypes, relating to the period between primary infection and hypnozoite activation \cite{lover2013quantifying}. In tropical regions, relapses occur frequently within a few weeks to a few months, whereas in temperate regions, relapses typically occur between six to 12 months after initial infection. This variation in relapse frequency relates to vector dynamics and the transmission potential of \pv. In temperate regions, slower-relapsing hypnozoites may allow the parasites to survive colder months when mosquitoes are less prevalent, whereas, in tropical regions, a faster relapsing frequency may allow the parasite to maximise its transmission potential \cite{battle2014geographical,hulden2011activation}. As relapses contribute to the majority of blood-stage infections, it is important to capture these relapse dynamics when modelling \pv~disease transmission. 

The methods of incorporating hypnozoites and their associated relapse dynamics vary across the \pv~modelling literature. Modellers have often adopted the approach of assuming a binary state (presence or absence) for hypnozoites harboured within an individual \cite{ishikawa2000prevalence,ishikawa2003mathematical,fujita2006modeling,robinson2015strategies,champagne2022using}. The \pv~hypnozoite reservoir (i.e. the number of hypnozoites) is known to be non-binary \cite{white2014modelling,white2016variation}. Due to this, more recent \pv~models have attempted to incorporate the complex hypnozoite dynamics and capture the impact of the hypnozoite reservoir on transmission dynamics \cite{white2014modelling,mehra2022hypnozoite,anwar2022multiscale,thesis_somya,anwar2023optimal}. 

The methods used to capture \pv~immunity also vary across the modelling literature. When individuals are first infected with malaria, they naturally develop some level of immunity. This immunity can be defined as the body's state of resistance to the infection, and, with each subsequent infection, this acquired immunity is enhanced \cite{bruce1985essential}. Modellers may consider different types of immunity when modelling \pv~transmission. This includes immunity against new infections, protection against severe malaria, anti-parasite immunity (i.e. the ability to control parasite density upon infection), clinical immunity (i.e. protection against clinical disease), and transmission-blocking immunity (i.e. immunity that reduces the probability of parasite transmission to mosquitoes) \cite{de1988modulation,de1991mathematical,doolan2009acquired}.

One of the primary reasons for modelling infectious disease transmission is to understand the potential impact of treatment strategies on incidence. In terms of \pv, a combination therapy, known as radical cure, is needed to target both the acute infection and the dormant hypnozoite reservoir \cite{wells2010targeting,taylor2019short,poespoprodjo2022supervised}. The two drugs include: (i) a drug that clears parasites from the blood (such as chloroquine or artemisinin-based combination therapy; and (ii) an 8-aminoquinoline drug that clears hypnozoites from the liver (such as primaquine or tafenoquine). Targeting the hypnozoite reservoir is crucial in controlling or eliminating \pv, as transmission can be re-established from the reactivation of hypnozoites \cite{white2014modelling}. Incorporating Glucose-6-phosphate dehydrogenase deficiency (G6PD) testing is recommended before administering primaquine or tafenoquine as these drugs can cause life-threatening haemolysis in individuals with G6PD deficiency, an enzymopathy affecting up to 30\% of individuals in malaria-endemic regions \cite{recht2018use}. 

Other interventions that have been modelled include vector control, mass drug administration (MDA), mass screening and testing (MSaT), and \pv~serological testing and treatment (\textit{Pv}SeroTAT). Vector control measures are recommended by the WHO in order to achieve elimination \cite{zuber2018multidrug}.
MDA is an effective intervention for controlling malaria and was advocated by the WHO in the 1950s to control malaria transmission \cite{hsiang2013mass}. MDA involves treating the entire population, or a well-defined sub-population, in a geographic location regardless of their infection status \cite{newby2015review,hsiang2013mass}, such that both individuals who are infected and non-infected are treated. In a radical cure MDA intervention, individuals are given artemisinin-based combination therapy to clear blood-stage parasites and primaquine (or tafenoquine) to clear hypnozoites. Due to the risks associated with radical cure treatment in G6PD–deficient individuals, mass administration of radical cure is not recommended by the WHO without first screening for G6PD deficiency \cite{world2021second,howes2012g6pd, watson2018implications}. Another strategy for reducing and eliminating malaria is MSaT. This involves identifying and treating infected individuals within a specific geographical location by mass testing of all individuals regardless of their symptom status \cite{singh2022mass}. MSaT is effective in reducing malaria transmission in areas with low to moderate malaria prevalence. However, its success depends on the availability of accurate diagnostic tools, effective antimalarial drugs, and strong community participation \cite{world2020world,kim2021systematic}. \textit{Pv}SeroTAT is a method for identifying individuals with recent blood-stage infections who are potential hypnozoite carriers by measuring antibodies and providing treatment with radical cure \cite{obadia2022developing}. This method can identify individuals likely harbouring a hypnozoite reservoir, therefore allowing targeted treatment. Mathematical modelling has been used to understand how these different intervention strategies may impact \pv~transmission \cite{ishikawa2003mathematical,white2018mathematical,aguas2012modeling}.

In this article, we synthesise the findings of a scoping review of existing mathematical models for population-level \pv~transmission to provide a comprehensive overview of the modelling frameworks and methods used to characterise \pv~dynamics. In Section \ref{method}, we provide the search and inclusion criteria. We discuss the search results in Section \ref{ch3/result} as per the categorical structure in Figure \ref{fig:review_schematic} before concluding remarks and open problems are presented in Section \ref{conclusion}.

\section{Methods} \label{method}
We conducted a literature search on the 21st of May 2023, using the databases PubMed and Google Scholar to capture all relevant studies using the search terms ``hypnozoite'', ``malaria'', ``vivax'', and ``mathematical model'' with Boolean operators. We screened the titles, abstracts and full text of articles for the following inclusion criteria: 
\begin{itemize}
    \item the paper either applied or described a mathematical model of population-level \pv~transmission dynamics, and;
    \item the mathematical model of \pv~incorporated hypnozoite dynamics, as this is a distinguishing feature of \pv~parasites compared to other \textit{Plasmodium} spp.
\end{itemize}

We excluded papers that:
\begin{itemize}
    \item were only concerned with the within-host dynamics of \pv. Although within-host models of \pv~dynamics are important for understanding \pv~transmission, they were not directly relevant to the aim of our study (i.e. to identify and compare mathematical models of population-level \pv~transmission). Papers that modelled dynamics at both the within-host and population level (i.e. multi-scale models) were included.
    \item only used or described mathematical models of \textit{Plasmodium} species other than \pv~(e.g. a mathematical model of \textit{P. falciparum} infectious disease dynamics). Models that accounted for both \pv~ and another \textit{Plasmodium} species were included.
    \item were currently only available as a preprint.
     
\end{itemize}

Search terms were conducted in English only, and only literature published in English were considered. No limitations regarding study location, publication status (e.g. accepted, but no preprints), publication type, or publication year were included. To enhance the probability of finding all relevant literature, we screened all references within the articles that met our inclusion and exclusion criteria. Articles were then downloaded to identify key components, which are discussed in Section \ref{ch3/result}.

We categorised models depending on whether they used a stochastic or deterministic approach, and whether they were compartmental or agent-based. Deterministic models have no random variation and typically utilise a compartmental structure within a population to form differential equations to track the rate of flows between compartments. Stochastic models incorporate random variation and are useful for questions and scenarios where small population numbers or extinction are involved. In terms of \pv~ infectious disease modelling, agent-based models explicitly model \pv~transmission dynamics at an individual-level, for example, modelling the interaction between humans and vectors and associating respective state variables and parameters to each individual and vector. In our review, we found that almost all stochastic models were also agent-based, so even though these features are not mutually exclusive, we categorised models as (i) deterministic compartmental models or (ii) stochastic agent-based models.

\section{Search results} \label{ch3/result}
\begin{figure}[h!]
\centering
  \includegraphics[width=.7\linewidth]{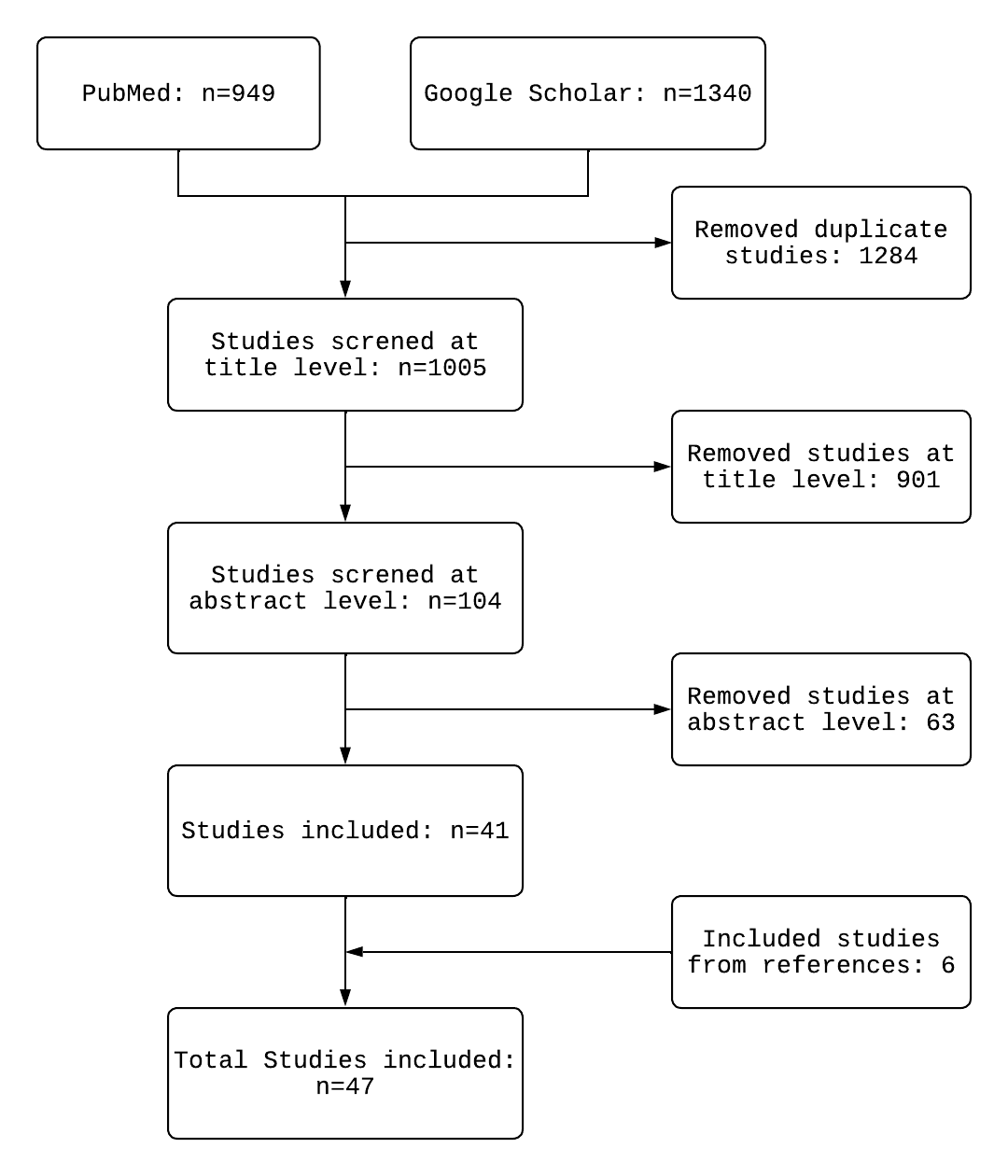} 
  \caption{\textit{Summary of the article selection process, illustrating papers included and excluded at each stage of the review process.}}
  \label{fig:review_process}
\end{figure}

The initial search yielded 2289 articles, which was reduced to 1005 unique articles after removing duplicates between the two databases. After screening at the title level, a further 901 studies were excluded as they did not fulfil the selection criteria in Section \ref{method}. After screening the abstracts, a further 63 studies were excluded due to either (i) no underlying mathematical model being described or (ii) the model was for \textit{P. falciparum} parasites only. Five additional studies were included from the selected studies' references that were not initially identified. A total of 47 studies were finally selected for review (see Figure \ref{fig:review_process} for a summary of the selection process). 

\begin{figure}[hptb!]
	% \vspace{1cm}
	% \begin{subfigure}[b]{1\textwidth}    
% 	\hspace{0.1cm}
	\begin{overpic}[width=.8\linewidth]{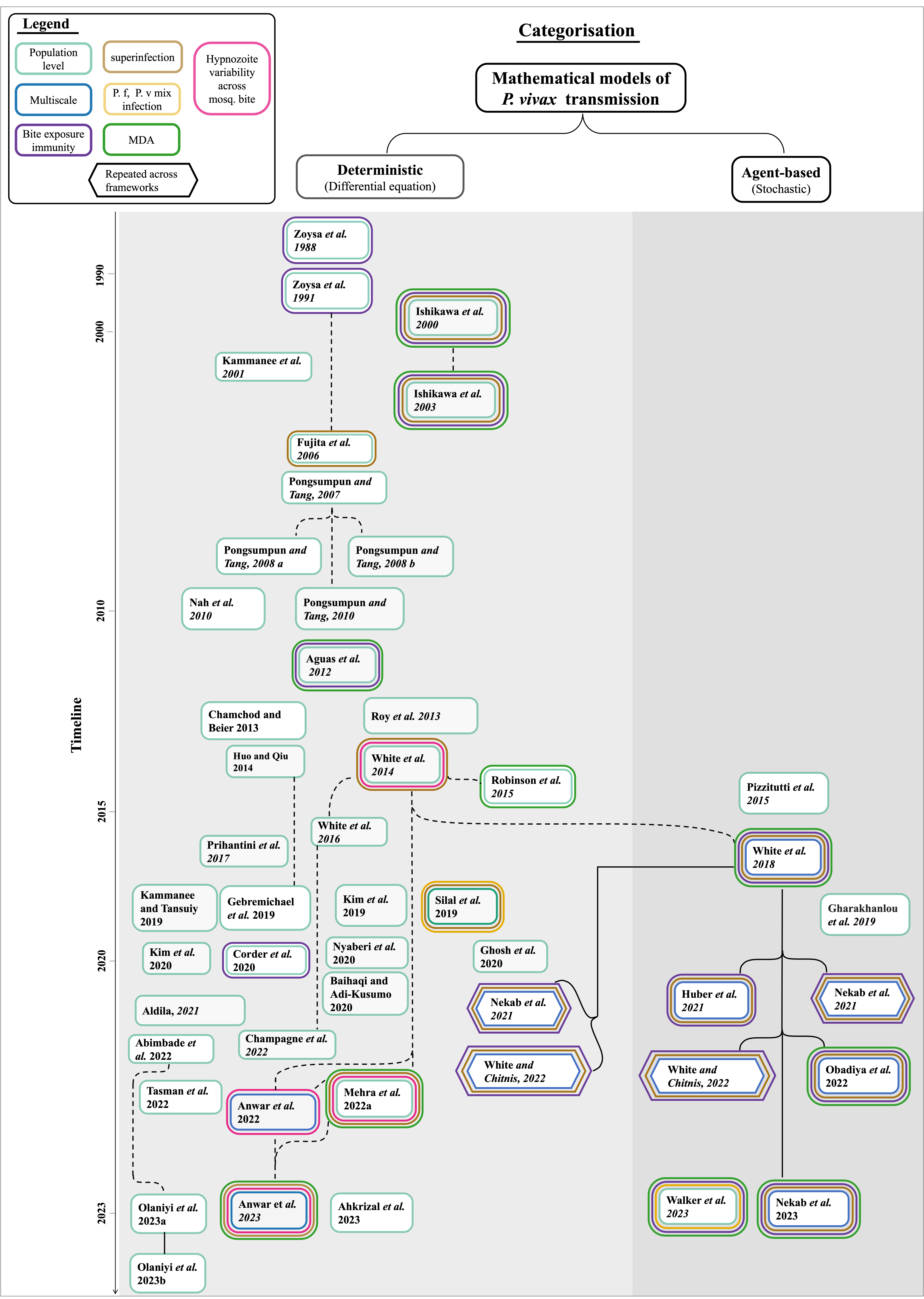}
	
 \put(25.5, 80.7){\fontsize{6}{6}\selectfont \cite{de1988modulation}}
	\put(25.5, 76.7){\fontsize{6}{6}\selectfont \cite{de1991mathematical}}
 
	\put(35.5, 74.63){\fontsize{6}{6}\selectfont \cite{ishikawa2000prevalence}}

 \put(20, 70.99){\fontsize{6}{6}\selectfont \cite{kammanee2001basic}}

	\put(35.5, 68.3){\fontsize{6}{6}\selectfont \cite{ishikawa2003mathematical}}
 
	\put(26, 64.67){\fontsize{6}{6}\selectfont \cite{fujita2006modeling}}

 \put(27.3, 61.65){\fontsize{6}{6}\selectfont \cite{pongsumpun2007transmission}}
 
 \put(22.5, 56.27){\fontsize{6}{6}\selectfont \cite{pongsumpun2008plasmodium}}

  \put(32.3, 56.2){\fontsize{6}{6}\selectfont \cite{pongsumpun2008mathematical}}

  \put(17.5, 52.2){\fontsize{6}{6}\selectfont \cite{nah2010dilution}}
  
   \put(28.3, 52.3){\fontsize{6}{6}\selectfont \cite{pongsumpun2010impact}}
   
 \put(27, 47.95){\fontsize{6}{6}\selectfont \cite{aguas2012modeling}}

  \put(21, 43.68){\fontsize{6}{6}\selectfont \cite{chamchod2013modeling}}

  \put(34.2, 44.45){\fontsize{6}{6}\selectfont \cite{roy2013potential}}

    \put(20.4, 40.58){\fontsize{6}{6}\selectfont \cite{huo2014stability}}

\put(31, 40.3){\fontsize{6}{6}\selectfont \cite{white2014modelling}}

         \put(41, 38.6){\fontsize{6}{6}\selectfont \cite{robinson2015strategies}}

         \put(60.5, 37.9){\fontsize{6}{6}\selectfont \cite{pizzitutti2015validated}}

         \put(26.6, 35.2){\fontsize{6}{6}\selectfont \cite{white2016variation}}

    \put(19.5, 33.66){\fontsize{6}{6}\selectfont \cite{prihantini2017stability}}

     \put(60.1, 33.2){\fontsize{6}{6}\selectfont \cite{white2018mathematical}}

     \put(14, 28.78){\fontsize{6}{6}\selectfont \cite{kammanee2019mathematical}}

     \put(21.5, 29.2){\fontsize{6}{6}\selectfont \cite{gebremichaelrelapse}}

         \put(28.6, 29.4){\fontsize{6}{6}\selectfont \cite{kim2019effects}}
         
    \put(35.5, 29.5){\fontsize{6}{6}\selectfont \cite{silal2019malaria}}

        \put(67.85, 28.6){\fontsize{6}{6}\selectfont \cite{gharakhanlou2019developing}}

   \put(13.9, 25.5){\fontsize{6}{6}\selectfont \cite{kim2020mathematical}}
   
   \put(21, 25.4){\fontsize{6}{6}\selectfont \cite{corder2020modelling}}
   
   \put(28.6, 25.85){\fontsize{6}{6}\selectfont \cite{nyaberi2020mathematical}}  

   \put(39.5, 25.6){\fontsize{6}{6}\selectfont \cite{ghosh2020mathematical}}  
   
   \put(29, 22.16){\fontsize{6}{6}\selectfont \cite{baihaqi2020modelling}} 
   
   \put(16.2, 22){\fontsize{6}{6}\selectfont \cite{aldila2021superinfection}}

      \put(40.2, 21.55){\fontsize{6}{6}\selectfont \cite{nekkab2021estimated}} 
      
      \put(55, 22.2){\fontsize{6}{6}\selectfont \cite{huber2021radical}} 

      \put(66.7, 22.35){\fontsize{6}{6}\selectfont \cite{nekkab2021estimated}} 

      \put(14.5, 18.5){\fontsize{6}{6}\selectfont \cite{abimbade2022recurrent}} 
      
      \put(22.5, 18.8){\fontsize{6}{6}\selectfont \cite{champagne2022using}} 

      \put(21.6, 13.47){\fontsize{6}{6}\selectfont \cite{anwar2022multiscale}} 
      
\put(29.4, 14.65){\fontsize{6}{6}\selectfont \cite{mehra2022hypnozoite}}  

\put(41.3, 17.85){\fontsize{6}{6}\selectfont \cite{white2022potential}}  

\put(55.5, 17.4){\fontsize{6}{6}\selectfont \cite{white2022potential}}

\put(66.2, 16.4){\fontsize{6}{6}\selectfont \cite{obadia2022developing}}  

\put(13.5, 14.65){\fontsize{6}{6}\selectfont \cite{tasman2022assessing}}

\put(53.5, 6.35){\fontsize{6}{6}\selectfont \cite{walker2023model}}

\put(62, 6.2){\fontsize{6}{6}\selectfont \cite{nekkab2023accelerating}}

\put(13.2, 5.8){\fontsize{6}{6}\selectfont \cite{olaniyi2023optimal}}

\put(21, 6.05){\fontsize{6}{6}\selectfont \cite{anwar2023optimal}}
 \put(29.4, 5.9){\fontsize{6}{6}\selectfont \cite{ahkrizal2023dynamics}}
 
 \put(13.5, 1.2){\fontsize{6}{6}\selectfont \cite{olaniyi2023efficiency}}
	\end{overpic}
	% \end{subfigure}
	\vspace{0.05cm}
  \caption{\textit{A summary of the 47 \pv~transmission models currently available in the literature (published as of May 21, 2023). Related models (either modified or motivated by) are connected with a dashed line. Similar/same models are connected with a solid line. The coloured boxes represent key features incorporated in the models (see legend). The hexagonal boxes with the same name represent that the model was also implemented in other frameworks. The timescale (non-linear) is shown on the left.}}
\label{fig:review_schematic}
\end{figure}

\subsection{Model frameworks}

In infectious disease dynamics, modelling frameworks typically involve a combination of mathematical models, statistical analyses, and computer simulations that aim to capture the complex dynamics of disease transmission. The Ross-Macdonald model \cite{macdonald1957epidemiology}, a compartmental model initially developed to describe malaria transmission dynamics, has been widely used as a modelling framework for \pv~transmission. This modelling approach has been adapted to investigate a range of vector-borne infectious diseases, and has helped inform public health policies and intervention strategies. 
The first mathematical model describing \pv~transmission was introduced by \mbox{---} to the best of our knowledge \mbox{---} Zoysa \textit{et al.} (1988) \cite{de1988modulation} in a Ross-Macdonald style modelling approach. Following this, many models have now been developed.

Out of the 47 studies identified that incorporated a \pv~transmission model, 37 (79\%) utilised a deterministic and differential equation (compartmental) framework \cite{de1988modulation,de1991mathematical,ishikawa2000prevalence,kammanee2001basic,ishikawa2003mathematical,fujita2006modeling,pongsumpun2007transmission,pongsumpun2008plasmodium,pongsumpun2008mathematical,pongsumpun2010impact,nah2010dilution,aguas2012modeling,chamchod2013modeling,roy2013potential,huo2014stability,white2014modelling,robinson2015strategies,white2016variation,prihantini2017stability,kammanee2019mathematical,gebremichaelrelapse,kim2019effects,silal2019malaria,kim2020mathematical,corder2020modelling,nyaberi2020mathematical,ghosh2020mathematical,baihaqi2020modelling,aldila2021superinfection,abimbade2022recurrent,tasman2022assessing,mehra2022hypnozoite,anwar2022multiscale,olaniyi2023optimal,anwar2023optimal,ahkrizal2023dynamics,olaniyi2023efficiency} and nine (19\%) used a stochastic and agent-based framework \cite{pizzitutti2015validated,white2018mathematical,gharakhanlou2019developing,huber2021radical,nekkab2021estimated,white2022potential,obadia2022developing,walker2023model,nekkab2023accelerating} (Figure \ref{fig:review_schematic}). Only one study (2\%) used both deterministic and stochastic frameworks \cite{robinson2015strategies} to model \pv~transmission. Robinson \etal(2015) \cite{robinson2015strategies} developed the model in a deterministic framework but implemented a stochastic version of the model as a continuous-time Markov chain. For simplicity, we categorise this model as deterministic in Figure \ref{fig:review_schematic}). Deterministic models are often the first choice amongst modellers due to their simplicity in comparison to stochastic models. Deterministic models are useful for understanding disease dynamics in large populations. Stochastic models provide more realistic and accurate representations of complex systems when dealing with small population sizes or low disease prevalence, as they can account for the randomness and variability observed in real life \cite{allen2000comparison,beran1994statistics}.

In contrast to the compartmental differential equation framework, agent-based models represent a system as a collection of individual agents that interact with each other based on a set of rules or behaviours \cite{volker2005individual,bonabeau2002agent}. The main difference between compartmental and agent-based modelling frameworks is that a compartmental model uses aggregate variables or compartments to represent the system, while agent-based models use individuals (agents) \cite{volker2005individual}. 
Out of the eight studies that used an agent-based model to capture the dynamics of \pv~transmission, only two studies modelled both the human and mosquito populations as agents \cite{pizzitutti2015validated,gharakhanlou2019developing}. The other agent-based models modelled the mosquito populations as a deterministic compartmental process, such that they combined ordinary differential equations for mosquitoes with an agent-based model for humans \cite{white2018mathematical, huber2021radical,nekkab2021estimated,obadia2022developing,white2022potential,walker2023model,nekkab2023accelerating}. Modelling mosquito dynamics as a deterministic process is an approximate strategy if the size of the mosquito population is very large and \pv~is not near elimination. In this case, the average behaviour of the stochastic dynamics agrees with those of a deterministic process \cite{mohd2022revisiting,bonabeau2002agent,yao2015law}. The actual behaviour of the system depends on the interactions between individuals and mosquitoes, instead of averages. Treating the mosquito compartment as a deterministic process means that modelling elimination is impossible, as there will always be some non-zero number of infectious mosquitoes remaining that can trigger an infection in humans again~\cite{walker2023model}.

Environmental features, ecology, and mosquito habitat locations were explicitly included when modelling malaria spread in the agent-based models that modelled both the human and mosquito population as agents \cite{pizzitutti2015validated,gharakhanlou2019developing}. The most recent agent-based models modelling \pv~dynamics \cite{ huber2021radical,nekkab2021estimated,obadia2022developing,white2022potential,nekkab2023accelerating} have evolved from a model introduced by White \etal(2018) \cite{white2018mathematical}. The White \etal(2018) \cite{white2018mathematical} model has been adapted to capture disease epidemiology in particular geographical settings \cite{nekkab2021estimated}, and to study the impact of different interventions (drugs or vaccination) \cite{huber2021radical,white2022potential,nekkab2023accelerating}. 

While agent-based models have many advantages, their use poses several challenges. One of the main challenges of agent-based models is the difficulty in parameterising and calibrating the model, given the large number of agents and their interactions \cite{crooks2008key,laubenbacher2020agent, colman1998complexity}. For example, despite being an agent-based model, parameterisation is done using an ordinary differential equation system that describes the process in several models \cite{white2018mathematical,huber2021radical,nekkab2021estimated,obadia2022developing}. Despite these challenges in parameterisation, agent-based models also often offer a more intuitive representation of epidemiological processes. The computational demands of agent-based models can be challenging \cite{samuelson2006agent}, although with improving computer technology, this has become less of a concern \cite{Darrin_Darrin_2021}.

\subsection{Population-level multiscale models}
Multiscale disease modelling incorporates at least two interacting scales and provides insights into disease dynamics across these scales that cannot be obtained from a single scale alone \cite{garabed2020multi}. Here we only focus on within-host population models as `multiscale models'. For \pv, multiscale modelling approaches can incorporate the complex hypnozoite dynamics and their relapse effects on onward disease transmission. Most models in the existing literature only capture the population-level impact of \pv~(boxes with a light lime green border in Figure \ref{fig:review_schematic}). Few models capture both within-host and population-level impacts (boxes with a strong blue border in Figure \ref{fig:review_schematic}) \cite{white2014modelling,white2018mathematical,nekkab2021estimated,huber2021radical,anwar2022multiscale,white2022potential,obadia2022developing}. The very first multiscale model for \pv~transmission was developed by White \textit{et al.} (2018) \cite{white2018mathematical}, and modelled the within-host hypnozoite dynamics using an agent-based model that considered heterogeneity in exposure to mosquito bites. This built on White \etal(2014) \cite{white2014modelling}, which was the first to develop a within-host model that captured the dynamics of \pv~hypnozoites.
This multiscale model considered the variability in the size of hypnozoite inoculum across each mosquito bite and was subsequently used to parameterise a separate transmission model that captured the entire structure of the hypnozoite reservoir \cite{white2018mathematical}. The White \etal(2014) \cite{white2014modelling} within-host model for temperate settings assumed collective dormancy. This means that the hypnozoites established by each mosquito bite progress through the dormancy states as a group or batch. This assumption may be biologically unrealistic due to the independence of individual hypnozoite activation and clearance dynamics within liver cells \cite{mehra2020activation}. The other within-host models that were adapted from White \etal(2018) \cite{white2018mathematical} applied the same assumption regarding batch hypnozoite behaviour \cite{ huber2021radical,nekkab2021estimated,obadia2022developing,white2022potential}.

Recent work by Mehra \textit{et al.} (2020) \cite{mehra2020activation}  relaxed the collective dormancy assumption. This enabled them to characterise the long-latency period of hypnozoite dynamics (a period of latency prior to hypnozoite activation) modelled (light purple bordered box in Figure \ref{fig:review_schematic}) in White \textit{et al.} (2014) \cite{white2014modelling} in analytical form. Later work by Mehra and colleagues embedded the activation-clearance model governing a single hypnozoite in an epidemiological framework \cite{mehra2022hypnozoite}. This framework accounts for successive mosquito bites, where each bite can simultaneously establish multiple hypnozoites \cite{mehra2021antibody,mehra2022hypnozoite}, and explores the epidemiological consequence of radical cure treatment on a single individual. Anwar \textit{et al.} (2022) \cite{anwar2022multiscale} have since developed a multiscale model motivated by White \etal(2014) \cite{white2014modelling} by embedding the framework of Mehra \textit{et al.} (2022) \cite{mehra2022hypnozoite} for short-latency hypnozoites (deriving the relapse rate by averaging the distribution of hypnozoite burden, which is dependent on the force of reinfection) into a simple population-level model that provides key insights into both within-host level and population level dynamics. The within-host and population models were coupled at each time step (thus producing a multiscale model) to incorporate key parameters that describe the hypnozoite dynamics. This multiscale model can provide the hypnozoite distributions within the population and, more importantly, reduces the infinite compartmental structure of White \textit{et al}. (2014) \cite{white2014modelling} into three compartments and relaxes the artificial truncation needed in White \textit{et al}. (2014) \cite{white2014modelling} for numerical simulation. Mehra \textit{et al.} (2022) \cite{thesis_somya} proposed an alternative approach, constructing a Markov population process to couple host and vector dynamics whilst accounting for (short-latency) hypnozoite accrual and superinfection as per the within-host framework proposed in Mehra \textit{et al.} (2022) \cite{mehra2022hypnozoite}. In the infinite population size limit, Mehra \textit{et al.} (2022) \cite{thesis_somya} recovered a functional law of large numbers for this Markov population process, comprising an infinite compartment deterministic model. This infinite compartment model was then reduced into a system of integrodifferential equations based on the expected prevalence of blood-stage infection derived at the within-host scale \cite{mehra2022hypnozoite}. This construction yielded population-level distributions of superinfection and hypnozoite burden, and has been generalised to allow for additional complexity, such as long-latency hypnozoites and immunity \cite{thesis_somya}.

\subsection{Hypnozoite dynamics and variation}
The eradication of \pv~is challenging due to the presence of the hypnozoite reservoir, which is undetectable and causes new infections long after the initial infection. In developing the first mathematical model for \pv,  Zoysa \textit{et al.} (1991) were also the first to model the effect of hypnozoite relapse on \pv~transmission \cite{de1991mathematical}. Since most \pv~blood-stage infections are due to the reactivation of hypnozoites rather than new primary infections, it is crucial that mathematical models incorporate the size of the hypnozoite reservoir \cite{baird2008real, betuela2012relapses,commons2019risk, commons2018effect,luxemburger1999treatment}. Zoysa \textit{et al.} (1991) \cite{de1991mathematical} assumed that the transmission dynamics could be accounted for by modelling a hypnozoite reservoir of size two (to account for up to two relapses). This assumption was later followed by Fujita \etal (2006) \cite{fujita2006modeling}. In reality, the average size of the hypnozoite reservoir is likely to be more than two in endemic settings, particularly those with high transmission intensity \cite{white2016variation}. Despite having the relapse characteristic that makes \pv~parasites unique, Aldila \etal (2021) \cite{aldila2021superinfection} did not incorporate relapses in their \pv~transmission model. In this model, individuals did not harbour hypnozoites when infected with \pv~and hence did not experience relapse after recovery from blood-stage infection.

Most \pv~transmission models consider the hypnozoite reservoir as a single compartment, rather than explicitly accounting for a variable number of hypnozoites in the reservoir \cite{ishikawa2000prevalence,kammanee2001basic,ishikawa2003mathematical,pongsumpun2007transmission,pongsumpun2008mathematical,pongsumpun2008plasmodium,nah2010dilution,pongsumpun2010impact,aguas2012modeling,chamchod2013modeling,roy2013potential,huo2014stability,robinson2015strategies,white2016variation,pizzitutti2015validated,prihantini2017stability,kammanee2019mathematical,gebremichaelrelapse,kim2019effects,silal2019malaria,gharakhanlou2019developing,kim2020mathematical,corder2020modelling,nyaberi2020mathematical,ghosh2020mathematical,baihaqi2020modelling,abimbade2022recurrent,tasman2022assessing,olaniyi2023optimal,ahkrizal2023dynamics}. Only a handful of models account for the variability in hypnozoite inoculation across mosquito bites (boxes with a bright pink border in Figure \ref{fig:review_schematic}) \cite{white2014modelling,mehra2022hypnozoite,anwar2022multiscale,anwar2023optimal}. If the size of the hypnozoite reservoir is modelled explicitly, the number of compartments in the model increases substantially. The very first model that accounted for the variation in hypnozoites across mosquito bites was introduced by \mbox{---} to the best of our knowledge \mbox{---} White \etal(2014) \cite{white2014modelling} for a short-latency strain (where hypnozoites can activate immediately after establishment). To account for the variation of hypnozoites across bites, White \textit{et al.} (2014) modelled a system with an infinite number of compartments to represent individuals with different numbers of hypnozoites. In practice, this is truncated at $2(L_{\text{max}}+1)$ ordinary differential equations (for human population only), where $L_{\text{max}}$ is the maximum number of hypnozoites considered. In their model, the hypnozoite reservoir within individuals increases with new infectious bites and decreases with both activation and death of hypnozoites. This infinite compartmental system makes the model very complex, particularly when other important structures must also be incorporated, such as individual heterogeneity in bite exposure. An agent-based model later developed by White \etal(2018) \cite{white2018mathematical}, and other models that utilise this agent-based model, consider variation in hypnozoites within individuals, but do not account for the variability in hypnozoites across mosquito bites \cite{huber2021radical,nekkab2021estimated,white2022potential,obadia2022developing}. Furthermore, instead of explicitly modelling hypnozoites independently, they impose the batch hypnozoite model. This assumption means that hypnozoites from a mosquito bite act as a batch, where they all reactivate simultaneously, causing relapse or dying at the same time. This reduces one batch of hypnozoites to a single set of dynamics, which is still truncated at a maximum of $k$ batches.

The multiscale model developed by Anwar \textit{et al.} (2022) \cite{anwar2022multiscale} accounted for the variation of hypnozoites dynamics across bites. Unlike the White \etal(2014) \cite{white2014modelling} model,  Anwar \textit{et al.} (2022)  only utilised three compartments at the population level by embedding the within-host model (short-latency) developed by Mehra \etal(2022) \cite{mehra2022hypnozoite} as a system of integrodifferential equations. This relaxes the artificial truncation for the maximum number of hypnozoites used within the White \etal(2014) \cite{white2014modelling} model. Under a constant force of reinfection, Anwar \textit{et al.} (2022) \cite{anwar2022multiscale} analytically proved that the multiscale model \cite{anwar2022multiscale} exhibits an identical steady-state hypnozoite distribution as the infinite ordinary differential equation model structure in White \etal(2014) \cite{white2014modelling}.  The advantage of the multiscale model by Anwar \textit{et al.} (2022) \cite{anwar2022multiscale} is that the population-level component is considerably simpler than the $2\big(L_{max}+1\big)$ ordinary differential equations of White \textit{et al.} (2014) \cite{white2014modelling}. The transmission models proposed by Mehra \textit{et al.} (2022) \cite{thesis_somya} likewise account for variation in hypnozoite batch sizes, with Mehra \textit{et al.} (2022) \cite{thesis_somya} additionally accommodating long-latency hypnozoite dynamics. The models of Mehra \textit{et al.} \cite{thesis_somya} are formulated as systems of integrodifferential equations, informed by the within-host framework of Mehra \textit{et al.} (2022) \cite{mehra2022hypnozoite}. The analyses of Anwar and Mehra \textit{et al.} \cite{anwar2022multiscale, thesis_somya} provided insights into hypnozoite dynamics (e.g. the average size of a hypnozoite reservoir within the population and the average relapse rate), in addition to disease dynamics.

\subsection{Superinfection} 
Superinfection with malaria is a common phenomenon and can be defined as when an individual has more than one blood-stage infection with the same malaria-causing parasite species at a given time  \cite{smith2010quantitative}. For \pf~malaria, when an infected individual (primary infection) receives a second infectious mosquito bite, they can become infected with two different parasite broods. In reference to \pv~malaria, individuals can harbour hypnozoites in the liver even after they recover from a primary infection. Therefore, relapsing hypnozoites can provide another pathway to superinfection for individuals infected with \pv~ \cite{portugal2011superinfection,smith2010quantitative}. 

When modelling \pv~dynamics, it is important to consider the impact of superinfection on recovery and transmission, as the abundance of mosquitoes and the contribution of hypnozoite activation can frequently trigger superinfection. Superinfection can potentially delay recovery from infection \cite{smith2012ross,dietz1974malaria}. Most of the literature that incorporates superinfection in \pv~transmission models (boxes with a brown border in Figure \ref{fig:review_schematic}) \cite{ishikawa2000prevalence,ishikawa2003mathematical,fujita2006modeling,white2014modelling,white2018mathematical,silal2019malaria,nekkab2021estimated,huber2021radical,white2022potential,olaniyi2023optimal} does so via the recovery rate \cite{ishikawa2000prevalence,ishikawa2003mathematical,fujita2006modeling,white2014modelling}. The superinfection phenomenon was first introduced into malaria models by Macdonald (1950) \cite{macdonald1950}, who assumed ``\textit{The existence of infection is no barrier to superinfection, so that two or more broods or organisms may flourish side by side}''. In the malaria modelling literature, it has been assumed that each brood could be cleared independently at a constant rate. Following this assumption, Dietz \etal (1974) \cite{dietz1974malaria} proposed a recovery rate under superinfection for \pf~malaria, derived at equilibrium under a constant force of reinfection. This form of the recovery rate was adopted in most studies that included superinfection via the recovery rate. This approach is straightforward when hypnozoites are integrated into the model as a binary state (i.e. an individual either has or does not have hypnozoites) \cite{ishikawa2000prevalence,ishikawa2003mathematical,fujita2006modeling}. Since White \etal(2014) \cite{white2014modelling} accounts for the variation of hypnozoites, they modified the recovery rate proposed by Dietz \etal(1974) \cite{dietz1974malaria} to account for the additional burden of hypnozoites; however, Mehra \etal(2022) \cite{thesis_somya} argued that this modified recovery rate does not hold in the presence of hypnozoite accrual. 

Generally, there are two approaches when incorporating superinfection: (i) using a corrected recovery rate that explicitly accounts for the history of past infections in the population and hypnozoite accrual dynamics \cite{nasell2013hybrid,thesis_somya,anwar2023optimal} and (ii) coupling the prevalence of blood-stage infection (derived under a within-host model that accounts for superinfection) directly to the proportion of infected mosquitoes \cite{thesis_somya}. The within-host model of Mehra \etal(2022) \cite{mehra2022hypnozoite} included superinfection, with each blood-stage infection (whether primary or relapse) being cleared independently, while the population-level model developed by Anwar \etal(2022) \cite{anwar2022multiscale}, which built on work of Mehra \etal(2022)\cite{mehra2022hypnozoite}, did not incorporate superinfection. A correction to account for superinfection, based on the recovery rate formulated by Nåsell \textit{et al.} (2013) \cite{nasell2013hybrid}, was proposed in Mehra \etal(2023) \cite{thesis_somya} and incorporated in later work by Anwar \etal(2023) \cite{anwar2023optimal}.

Superinfection was incorporated in later work, where it was assumed that different batches of hypnozoites originated from different mosquito bites \cite{white2018mathematical,huber2021radical,white2022potential,obadia2022developing}. Silal \etal(2019) \cite{silal2019malaria} assumed that superinfection increased the severity of the disease. That is, individuals will transition from lower to higher severity classes with a certain probability due to multiple infections. The only other study incorporating a superinfection-like phenomenon was Aldila \etal(2021), who modelled \pv~and \pf~co-infection and assumed that \pv~dominates \pf~\cite{aldila2021superinfection}, which does not closely resemble the definition of superinfection. This study assumed that if an individual was currently infected with \pf, they would become infected with \pv~if they received an infectious bite from a mosquito that was infected with \pv. The assumption that \pv~parasites dominate \pf~results in the individual being infected with only \pv, which is not supported by the empirical biological evidence that shows that the parasitaemic load is much higher for \pf~\cite{battle2021global}. Accordingly, it may not be reasonable to consider this to be a valid model of superinfections.

\subsection{\pv~and \pf~ co-infection}

Within the Asia-Pacific region, the horn of Africa, and South America, both \pv~and \pf~parasites are common \cite{silal2019malaria,walker2023model}. For example, in 2019 in Cambodia, co-infection with both \pv~and \pf~accounted for about $17\%$ of malaria cases \cite{chhim2021malaria}. In co-endemic regions, \pf~infections are often followed by \pv~infection, giving rise to the hypothesis that \pf~infections trigger \pv~hypnozoite activation \cite{snounou2004co,white2011determinants,lin2011plasmodium,silal2019malaria}. The high risk of \pv~parasitaemia after \pf~infection is possibly related to reactivation of hypnozoites \cite{taylor_resolving_2019,hossain_risk_2020,commons_risk_2019}. Hypnozoites may be activated when \pf~parasites have been introduced into the body \cite{shanks2013activation} or when the individual is exposed to \textit{Anopheles} specific proteins \cite{hulden2011activation}. This increased risk of \pv~blood-stage infection following a \pf~infection could alternatively be explained by spatial or demographic heterogeneity in exposure and thus infection risk. Individuals either living in areas where both \pv~and \pf~are highly prevalent or those that engage in an activity bringing them in frequent contact with infected mosquitoes (e.g. forest work) are more likely to be exposed to both parasites than the average person. Having a \textit{P. falciparum} episode indicates the person has recently been exposed to infectious mosquito bites and is thus likely to have hypnozoites from previous exposure events (that may be triggered or activated spontaneously) or acquire a new primary \textit{P. vivax} infections following recovery from \textit{P. falciparum} infection \cite{amratia2019characterizing,haque2010spatial,hofmann2017complex}. The lack of diagnostics to differentiate primary infections and relapses further complicates determining when an individual is infected with \pv~hypnozoites. This makes it challenging to disentangle whether \pf~infections cause relapses through the reactivation of hypnozoites.

It is also not yet clearly understood how \pv~and \pf~interact, if they compete within the host or if one species causes some, if any, protection against the other \cite{muh2020cross,cox2008knowlesi}.  A systematic review and meta-analysis showed that mixed infections (\pf~and \pv) can often cause a high rate of severe infection regardless of infection order \cite{kotepui2020plasmodium}. This evidence was in contrast to a previous study which suggested that severe mixed infections were more likely to happen when \pv~infection occurred on top of an existing \pf~infection (i.e. superinfection), whereas the reverse scenario, \pf~infection on top of an existing \pv~infection, were more likely to result in a lower risk of severe malaria \cite{mohapatra2012profile}. Furthermore, there is likely ascertainment bias associated with mixed infections in areas with co-circulating parasite strains, as efforts might be biased towards \pf~detection \cite{stresman2020association}. This is likely to be particularly common during episodes of clinical malaria when parasitaemia of one species greatly exceeds the other, and the innate host immune response may suppress both infections. Gaining a better understanding of these cross-species interactions and adjusting accounting for this co-existing phenomenon in the co-endemic region will require multi-species transmission models. Only a handful of mathematical models included both these \textit{Plasmodium} species \cite{pongsumpun2008mathematical,pongsumpun2010impact,aguas2012modeling,pizzitutti2015validated,silal2019malaria,aldila2021superinfection}. While both \pv~and \pf~species are included in a single model by Aldila \etal(2021) \cite{aldila2021superinfection}, this model did not account for \pv~ relapses. Five studies included both species but used two independent models for each species, which did not allow for interactions between species \cite{pongsumpun2008mathematical,pongsumpun2010impact,aguas2012modeling,pizzitutti2015validated,aldila2021superinfection}. 

Whether it is important to model species interaction depends on the particular geographical setting. If both parasites are co-endemic in a setting, and the research question being considered relates to both species, then it may be important to use a model that can capture the interactions between the parasite species \cite{snounou2004co,silal2019malaria,walker2023model}. To the best of our knowledge, the first model that accounts for the interaction between both species was developed by Silal \etal(2019) \cite{silal2019malaria}. In this study, a separate model (deterministic, meta-population) for both species was proposed, and these two models were entangled at each time step to incorporate interactions between the species, including treatment, triggering, and masking (non-\pf~infections are misdiagnosed as \pf).  Following this work, the first agent-based model transmission model accounting for both \pv~and \pf~infections and treatment was developed by Walker \etal(2023) \cite{walker2023model}. This model had reduced complexity compared with Silal \textit{et al.}'s (2019) co-infection model, but used many of the same parameter values \cite{silal2019malaria} (co-infection models shown with a vivid orange bordered box in Figure \ref{fig:review_schematic}). 

\subsection{Immunity}
Immunity against disease acquired through infection is usually referred to as adaptive immunity, and the primary function of adaptive immunity is to destroy foreign pathogens \cite{mayer2010microbiology,chaffey2003alberts}. Naturally acquired immunity to malaria is characterised by relatively rapid acquisition of immunity against severe disease and a more gradual establishment of immunity against uncomplicated malaria, while sterile immunity against infections is never achieved \cite{barua2019impact,gruner2007sterile,mueller2013natural,lopez2017known}. In co-endemic areas, clinical immunity to \textit{P. vivax} is more rapidly acquired than that due to \textit{P. falciparum} \cite{mueller2013natural}. 

How immunity is accounted for in mathematical models of malaria varies since different models consider different types of immunity. For example, immunity against new infections, immunity against severe malaria, anti-parasite immunity (i.e. the ability to control parasite density upon infection), clinical immunity (i.e. protection against clinical disease), and transmission-blocking immunity (i.e. reducing the probability of parasite transmission to mosquitoes). Immunity against new infections and severe malaria is assumed to be acquired through infection. This reduces the probability of reinfection from an infectious mosquito bite and has been modelled using up to two immunity levels \cite{de1988modulation,de1991mathematical}. This type of immunity is assumed to be boosted by infection \cite{barua2019impact}. Acquiring some partial immunity (i.e. some degree of protection against malaria) following infection that wanes over time, is most common among published models \cite{kammanee2001basic,pongsumpun2007transmission,pongsumpun2008mathematical,pongsumpun2008plasmodium,pongsumpun2010impact,aguas2012modeling,chamchod2013modeling,kammanee2019mathematical,gebremichaelrelapse,gharakhanlou2019developing,tasman2022assessing}. Some assumptions regarding immunity include that, if treated, individuals acquire some level of immunity that reduces the probability of reinfection (i.e. gain immunity against new infection) and that this wanes over time \cite{roy2013potential,white2014modelling}. The assumption regarding permanent immunity against malaria is not considered valid, as immunity often wanes rapidly when immune adults leave malaria-endemic regions \cite{langhorne2008immunity}. Despite this, one model assumed that recovered individuals become permanently immune to \pv~\cite{aldila2021superinfection}. Another study assumed that only a fixed proportion of individuals are immune against \pv~rather than explicitly incorporating immunity into the model \cite{pizzitutti2015validated}. Strategies for incorporating immunity into \pv~transmission models thus widely vary, where some assumptions are more realistic and appropriate than others.

Individuals who have not previously experienced malaria infection almost invariably become infected when first exposed to infectious mosquito bites, as immunity against malaria has not yet developed \cite{langhorne2008immunity}. Repeated exposure to infectious bites will still likely result in infection, though these individuals may be protected against severe malaria or death \cite{langhorne2008immunity}. Silal \etal(2019) \cite{silal2019malaria} applied the opposite assumption and assumed that repeated exposure to infectious bites would likely result in severe infection \cite{silal2019malaria}. With increasing exposure, naturally acquired immunity will also give some level of protection against symptomatic malaria. Adults living in endemic areas are more likely to have developed protective immunity compared to children due to repeated exposure over their lifetime. Adults living in endemic areas are likely to have experienced substantially more infectious mosquito bites compared to children due to age (and therefore lengthened opportunity to acquire infectious mosquito bites), greater skin surface area, and more time spent outside in environments with a higher prevalence of mosquitoes \cite{port1980relationship,carnevale1978aggressiveness,white2018mathematical}. 

Immunity should be considered when using mathematical models to capture underlying disease dynamics. The assumption regarding immunity varies among models (boxes with a purple border in Figure \ref{fig:review_schematic}). The only model that explicitly accounts for the acquisition of immunity that increases with new bites was developed by White \etal (2018) \cite{white2018mathematical}. The assumption in regards to both anti-parasite immunity (ability to reduce parasite density upon infection) and clinical immunity (protection against clinical disease) depends on age and exposure to mosquito bites which is modelled using partial differential equations \cite{white2018mathematical}. They also assumed that children acquired immunity through their birth parent's immunity, which then decayed exponentially from birth. Models that were adapted from White \etal(2018) \cite{white2018mathematical} also allow for the acquisition of immunity \cite{huber2021radical,nekkab2021estimated,white2022potential,obadia2022developing}. However, the immunity acquired from a primary infection may protect more strongly against relapses (which are genetically related to the primary infection) than against a new, genetically distinct primary infection. That is, hypnozoites established from an infectious bite, when reactivated, may be less likely to cause clinical infection. This is because the parasites could be genetically identical or related, which could elicit a more protective immune response due to familiarity with the primary infection \cite{white2011determinants,joyner2019humoral}. Thus, relapses from the same batch of hypnozoites may only cause asymptomatic infections. Despite this, no models to date have fully accounted for the relationship between relapse and immunity. Model assumptions regarding the acquisition of immunity may be too simple to capture the true underlying biology and dynamics.

\subsection{Effect of interventions for malaria control}

In most of the models included in this review, it was assumed that treatment would be targeted towards infected individuals \cite{fujita2006modeling,chamchod2013modeling,roy2013potential,huo2014stability,white2014modelling,pizzitutti2015validated,gebremichaelrelapse,silal2019malaria,kim2012genotyping,nekkab2021estimated,huber2021radical,tasman2022assessing,olaniyi2023optimal}, but a range of other interventions can contribute to malaria control. For example, ten studies (21\%) evaluated the effect of MDA on disease transmission, despite few national control programs considering MDA for \pv~control \cite{ishikawa2000prevalence,ishikawa2003mathematical,aguas2012modeling,robinson2015strategies,white2018mathematical,mehra2022hypnozoite,obadia2022developing,anwar2023optimal,walker2023model,nekkab2023accelerating} (boxes with a dark green border in Figure \ref{fig:review_schematic}). Since MDA is recommended as an important tool to reduce asymptomatic \pf~infection, it is also likely to be of great importance for \pv~elimination \cite{robinson2015strategies,grueninger2013transitioning,okell2011potential}. One study examined the effect of multiple MDAs and MSaTs (up to two rounds) with different drug combinations (blood-stage drug only, blood-stage drug and primaquine, or blood-stage drug and tafenoquine), finding that MDA with tafenoquine following G6PD screening could significantly reduce transmission compared to MSaT, given that no tools were available at the time to identify individuals with hypnozoites \cite{robinson2015strategies}. The effect of long-lasting insecticide nets along with MDA was studied using an agent-based model in Papua New Guinea, where the model predicted that MDA could reduce \pv~transmission by between 58\% and 86\% \cite{white2018mathematical}. The same agent-based model was later used to investigate the effect of multiple treatment strategies, including MDA, MSaT with light microscopy detection of blood-stage parasitemia, and \pv~serological test and treatment (PvSeroTAT) \cite{obadia2022developing,nekkab2023accelerating}, as well as the effect of chloroquine and primaquine with vector control \cite{nekkab2021estimated}, and the potential effect of three different types of vaccines that target different stages of the \pv~life cycle \cite{white2022potential} in different geographical settings. The mixed-species agent-based model \cite{walker2023model} was used to investigate different treatment scenarios, including current practice, accelerated radical cure, and unified radical cure provided with and without MDA (radical cure was with 14 days of primaquine and a G6PD test while the MDA was with blood-stage treatments only).

 The only within-host model that accounted for the effect of MDA on each of the hypnozoites and infections was proposed by Mehra \etal(2022) \cite{mehra2022hypnozoite}. This model provided base analytical expressions for the effect of multiple rounds of MDA on hypnozoite dynamics and provided the epidemiological impact of one round of MDA on a single individual. Anwar \etal(2023) recently embedded Mehra \textit{et al.}'s work \cite{mehra2022hypnozoite} and extended the model to study the effect of multiple MDA rounds (up to $N$ rounds) on both within-host and population-level \cite{anwar2023optimal}. To the best of our knowledge, no other multiscale model has been developed that explicitly accounts for the effect of multiple rounds of MDA. Anwar \etal(2023) \cite{anwar2023optimal}
 also provided optimal intervals if multiple MDA rounds were under consideration.

\section{Open questions and conclusion}\label{conclusion}
  
Mathematical modelling is a powerful tool for understanding, analyzing, and predicting complex real-world phenomena, as well as simulating different scenarios, testing hypotheses, and making informed decisions based on the results. Mathematical models have proven useful to characterise \pv~transmission in different parts of the world and provide insights into the effect of different strategies to achieve elimination, including treatment, vaccination, and vector control. In this work, we provided a review of the existing mathematical models that capture \pv~disease progression and transmission. \pv~transmission dynamics are particularly challenging to model given the difficulties discerning relapses from reinfections and recrudescences. The choice of transmission model framework comes down to the research question at hand.

While mathematical models can provide key insights without the expense of large trials or epidemiological studies, it is important to recognize that mathematical models are not perfect representations of reality and are always subject to limitations, uncertainties, and assumptions. Therefore, using mathematical models in conjunction with empirical data, expert knowledge, and critical thinking is essential to obtain meaningful and reliable results. 

Across the different approaches of mathematical modelling of \pv, there were varying assumptions regarding parasite dynamics and acquisition of immunity. Some models were motivated to capture realistic biological aspects of the parasite \cite{de1988modulation,white2014modelling,anwar2022multiscale}, or epidemiological and public health aspects \cite{de1991mathematical,ishikawa2000prevalence,ishikawa2003mathematical,fujita2006modeling,aguas2012modeling,chamchod2013modeling,roy2013potential,robinson2015strategies,pizzitutti2015validated,white2016variation,white2018mathematical,silal2019malaria,gharakhanlou2019developing,corder2020modelling,nekkab2021estimated,huber2021radical,obadia2022developing,anwar2023optimal,tasman2022assessing,white2022potential,mehra2022hypnozoite,walker2023model}, whereas some models were motivated to construct a novel or extended mathematical model of \pv~dynamics, i.e., focusing on the mathematical aspects of \pv~dynamics \cite{kammanee2001basic,pongsumpun2007transmission,pongsumpun2008mathematical,pongsumpun2008plasmodium,pongsumpun2010impact,nah2010dilution,huo2014stability,prihantini2017stability,kammanee2019mathematical,gebremichaelrelapse,kim2019effects,kim2020mathematical,nyaberi2020mathematical,ghosh2020mathematical,aldila2021superinfection,abimbade2022recurrent,olaniyi2023optimal}. As the dynamics of these type of models are well established, we argue that more importance should be placed on using these models to address the current hurdles and setbacks in achieving \pv~elimination. For example, the effect of new drugs, emerging drug resistance, and the potential effect of vaccination (when it becomes available). Modelling different scenarios with the available tools under the current recommendations is crucial to inform decision-making regarding malaria elimination. Furthermore, given that some of the biological aspects of \pv~are well understood, we argue that researchers should shift their focus to modelling these important aspects. 

The spatial distribution of \pv~transmission is heterogeneous, and the number of hypnozoites that an individual harbours might vary significantly; this contributes directly to the risk of hypnozoite reactivation and \pv~relapse \cite{popovici2015challenges,stadler2022population}. This heterogeneity can be partially captured by modelling individuals' movement using metapopulations and including parasite movement between different sub-populations \cite{white2018mathematical,nekkab2021estimated}. However, none of the current models explicitly consider this spatial heterogeneity. Given the high degree of heterogeneity of \pv~risk in almost all populations, future model development should address this.

As more than 80\% of \pv~infections may be due to relapse, and multiple hypnozoites can be established from each infectious bite, modelling the dynamics of hypnozoite variation and activation is crucial \cite{mehra2022hypnozoite,anwar2022multiscale,robinson2015strategies}. Another important aspect that requires more detailed attention is the interaction between multiple species of \textit{Plasmodium}, particularly in areas where \pf~and \pv~are co-endemic (Asia, the Horn of Africa, and the Americas). Studies show that there is a high risk of \pv~parasitaemia after \pf~infection that is possibly related to reactivation of hypnozoites \cite{taylor_resolving_2019,hossain_risk_2020,commons_risk_2019}. This is in line with the hypothesis that \pf~infection might trigger underlying \pv~infection \cite{snounou2004co,white2011determinants,lin2011plasmodium,silal2019malaria}. Hence, we argue that this hypothetical triggering phenomenon should be investigated when modelling \pf~and \pv~interactions.

Future \pv~modelling efforts should also account for superinfections. Where mosquito abundance is high, transmission intensity is also likely to be high if malaria parasites are present \cite{portugal2011superinfection,bashar2014seasonal,reimer2016malaria,keven2022vector}. In these scenarios, infected individuals are likely to experience multiple episodes of infection at once (i.e. superinfection). Superinfection can significantly delay recovery time, leaving ample opportunity for onward transmission from the infected individual to susceptible mosquitoes. \pv~models should hence account for the transmission dynamics associated with superinfection.  Immunity against \pv~ strongly correlates to past exposure; therefore, focus should also be placed on modelling the acquisition (and waning) of immunity related to superinfection, as multiple concurrent exposures may boost immunity more than singular exposures \cite{thesis_somya,white2018mathematical}. Furthermore, as parasites from relapse are either genetically identical or related to a previous primary infection, they are more efficiently targeted by naturally acquired immune responses previously developed from the primary infection than further, genetically unrelated primary infections. As a consequence, relapses are less likely to be associated with (severe) clinical symptoms.  \cite{white2011determinants,joyner2019humoral}. This interplay between immunity and relapse has not been fully addressed in any models developed to date. Given these important biological aspects, we suggest that future modelling should focus on developing the above-mentioned key areas: (i) spatial heterogeneity in exposure risk, (ii)  accumulation of hypnozoites variation, (iii) \pf~and \pv~interactions, (iv) acquisition of immunity, and (v) recovery under superinfection. Different modelling communities have recently started focusing on these areas recently, for example, modelling hypnozoite dynamics \cite{mehra2020activation,mehra2022hypnozoite}, multispecies interactions (\pf~and \pv) \cite{silal2019malaria,walker2023model}, bite exposure immunity \cite{white2018mathematical} and superinfection \cite{mehra2022hypnozoite,thesis_somya,anwar2023optimal}. No model currently includes all of the above factors that play a role in \pv~transmission due to the complexity the resulting model would have, and not all of the factors may need to be modelled to answer the research questions at hand. Therefore, when developing models to explore \pv~disease progression with a focus on answering specific research questions, mathematical epidemiologists and modellers should consider relevant aspects within the context of existing recommendations. 

To address the outstanding research questions identified here, a suitably skilled interdisciplinary team is required. We hope that this review can contribute to developing the common language needed for communication between different scientists by highlighting the progress of \pv~transmission models to date.

\section*{Funding}
L. Smith is supported by the National Health and Medical Research Council (NHMRC) (GNT2016726) and the Department of Foreign Affairs and Trade Australia through the project Strengthening Preparedness in the Asia-Pacific Region through Knowledge (SPARK). A. Devine’s research is supported through the NHMRC (2019152). E. Conway, and I. Mueller's research are supported by the NHMRC (GNT2016726) and the Department of Foreign Affairs and Trade Australia through the project Strengthening Preparedness in the Asia-Pacific Region through Knowledge (SPARK). J.M. McCaw’s research is supported by the Australian Research Council (DP210101920) and the NHMRC Australian Centre of Research Excellence in Malaria Elimination (ACREME) (APP1134989). J.A. Flegg’s research is supported by the Australian Research 
Council (DP200100747, FT210100034) and the NHMRC (APP2019093). The contents of the published material are solely the responsibility of the individual authors and do not reflect the views of NHMRC.

\printbibliography[heading=bibintoc] %was in template

@article{dietz1974malaria,
  title={A malaria model tested in the African savannah},
  author={Dietz, K and Molineaux, L and Thomas, A},
  journal={Bulletin of the World Health Organization},
  volume={50},
  number={3-4},
  pages={347},
  year={1974},
  publisher={World Health Organization}
}

@article{de1988modulation,
  title={Modulation of human malaria transmission by anti-gamete transmission blocking immunity},
  author={de Zoysa, Arjuna PK and Herath, Pushpa RJ and Abhayawardana, TA and Padmalal, UKGK and Mendis, Kamini N},
  journal={Transactions of the Royal Society of Tropical Medicine and Hygiene},
  volume={82},
  number={4},
  pages={548--553},
  year={1988},
  publisher={Elsevier}
}

@article{de1991mathematical,
  title={A mathematical model for Plasmodium vivax malaria transmission: estimation of the impact of transmission-blocking immunity in an endemic area.},
  author={de Zoysa, AP and Mendis, C and Gamage-Mendis, AC and Weerasinghe, S and Herath, PR and Mendis, Kamini N},
  journal={Bulletin of the World Health Organization},
  volume={69},
  number={6},
  pages={725},
  year={1991},
  publisher={World Health Organization}
}

@article{ishikawa2000prevalence,
  title={The prevalence of Plasmodium vivax in Vanuatu Islands: Computer simulation of malaria control trails},
  author={Ishikawa, Hirofumi and Ishii, Akira and Kaneko, Akira},
  journal={Journal of the Faculty of Environmental Science and Technology},
  volume={5},
  number={1},
  pages={1--6},
  year={2000},
  publisher={Okayama University}
}

@article{kammanee2001basic,
  title={Basic reproduction number for the transmission of Plasmodium vivax malaria.},
  author={Kammanee, A and Kanyamee, N and Tang, IM},
  journal={The Southeast Asian Journal of Tropical Medicine and Public Health},
  volume={32},
  number={4},
  pages={702--706},
  year={2001}
}

@article{ishikawa2003mathematical,
  title={A mathematical model for the transmission of Plasmodium vivax malaria},
  author={Ishikawa, Hirofumi and Ishii, Akira and Nagai, Nobuhiko and Ohmae, Hiroshi and Harada, Masakazu and Suguri, Setsuo and Leafasia, Judson},
  journal={Parasitology International},
  volume={52},
  number={1},
  pages={81--93},
  year={2003},
  publisher={Elsevier}
}

@article{fujita2006modeling,
  title={Modeling of re-emerging Plasmodium vivax in the Northern Area of the Republic of Korea Based on a Mathematical Model},
  author={Fujita, Kazutoshi and Tian Tian, Chen and Nishina, Tomohiko and Ishikawa, Hirofumi},
  journal={Journal of the Faculty of Environmental Science and Technology},
  volume={11},
  number={1},
  pages={1--7},
  year={2006},
  publisher={Okayama University}
}

@inproceedings{pongsumpun2007transmission,
  title={Transmission model for Plasmodium vivax malaria},
  author={Pongsumpun, Puntani and Tang, I-Ming},
  booktitle={Proceedings of the 3rd WSEAS/IASME International Conference on Dynamical Systems and Control},
  pages={276--281},
  year={2007}
}

@article{pongsumpun2008plasmodium,
  title={Plasmodium Vivax Malaria Transmission in a Network of Villages},
  author={Pongsumpun, P and Tang, IM},
  journal={Eng and Tech},
  pages={333--337},
  year={2008},
  publisher={Citeseer}
}

@article{pongsumpun2008mathematical,
  title={Mathematical model for the transmission of P. falciparum and P. vivax malaria along the Thai-Myanmar border},
  author={Pongsumpun, Puntani and Tang, I-Ming},
  journal={International Journal of Biological and Medical Sciences},
  volume={3},
  number={3},
  year={2008}
}

@article{pongsumpun2010impact,
  title={Impact of cross-border migration on disease epidemics: case of the P. falciparum and P. vivax malaria epidemic along the Thai-Myanmar border},
  author={Pongsumpun, Puntani and Tang, I-Ming},
  journal={Journal of Biological Systems},
  volume={18},
  number={01},
  pages={55--73},
  year={2010},
  publisher={World Scientific}
}

@article{nah2010dilution,
  title={The dilution effect of the domestic animal population on the transmission of P. vivax malaria},
  author={Nah, Kyeongah and Kim, Yongkuk and Lee, Jung Min},
  journal={Journal of Theoretical Biology},
  volume={266},
  number={2},
  pages={299--306},
  year={2010},
  publisher={Elsevier}
}

@article{aguas2012modeling,
  title={Modeling the effects of relapse in the transmission dynamics of malaria parasites},
  author={{\'A}guas, Ricardo and Ferreira, Marcelo U and Gomes, M Gabriela M},
  journal={Journal of Parasitology Research},
  volume={2012},
  year={2012},
  publisher={Hindawi}
}

@article{chamchod2013modeling,
  title={Modeling Plasmodium vivax: relapses, treatment, seasonality, and G6PD deficiency},
  author={Chamchod, Farida and Beier, John C},
  journal={Journal of Theoretical Biology},
  volume={316},
  pages={25--34},
  year={2013},
  publisher={Elsevier}
}

@article{roy2013potential,
  title={The potential elimination of Plasmodium vivax malaria by relapse treatment: insights from a transmission model and surveillance data from NW India},
  author={Roy, Manojit and Bouma, Menno J and Ionides, Edward L and Dhiman, Ramesh C and Pascual, Mercedes},
  journal={PLoS Neglected Tropical Diseases},
  volume={7},
  number={1},
  year={2013},
  publisher={Public Library of Science}
}

@inproceedings{huo2014stability,
  title={Stability of a mathematical model of malaria transmission with relapse},
  author={Huo, Hai-Feng and Qiu, Guang-Ming},
  booktitle={Abstract and Applied Analysis},
  volume={2014},
  year={2014},
  organization={Hindawi}
}

@article{white2014modelling,
  title={Modelling the contribution of the hypnozoite reservoir to Plasmodium vivax transmission},
  author={White, Michael T and Karl, Stephan and Battle, Katherine E and Hay, Simon I and Mueller, Ivo and Ghani, Azra C},
  journal={Elife},
  volume={3},
  pages={e04692},
  year={2014},
  publisher={eLife Sciences Publications Limited}
}

@article{robinson2015strategies,
  title={Strategies for understanding and reducing the Plasmodium vivax and Plasmodium ovale hypnozoite reservoir in Papua New Guinean children: a randomised placebo-controlled trial and mathematical model},
  author={Robinson, Leanne J and Wampfler, Rahel and Betuela, Inoni and Karl, Stephan and White, Michael T and Suen, Connie SN Li Wai and Hofmann, Natalie E and Kinboro, Benson and Waltmann, Andreea and Brewster, Jessica and others},
  journal={PLoS Med},
  volume={12},
  number={10},
  pages={e1001891},
  year={2015},
  publisher={Public Library of Science}
}

@article{white2016variation,
  title={Variation in relapse frequency and the transmission potential of Plasmodium vivax malaria},
  author={White, Michael T and Shirreff, George and Karl, Stephan and Ghani, Azra C and Mueller, Ivo},
  journal={Proceedings of the Royal Society B: Biological Sciences},
  volume={283},
  number={1827},
  pages={20160048},
  year={2016},
  publisher={The Royal Society}
}

@article{prihantini2017stability,
  title={Stability analysis model of Plasmodium vivax from Anopheles in human infection using SIDR population compartment with treatment RTS-S/AS01 vaccine in Yogyakarta},
  author={Prihantini and Pratama, Yoga Bagas and Muhammad, Ammar and others},
  journal={EDUCATUM Journal of Science, Mathematics and Technology},
  volume={4},
  number={2},
  pages={43--50},
  year={2017}
}

@article{kammanee2019mathematical,
  title={A mathematical model of transmission of Plasmodium vivax malaria with a constant time delay from infection to infectious},
  author={Kammanee, Athassawat and Tansuiy, Orawan},
  journal={Communications of the Korean Mathematical Society},
  volume={34},
  number={2},
  pages={685--699},
  year={2019},
  publisher={Korean Mathematical Society}
}

@article{gebremichaelrelapse,
  title={Relapse Effect on the Dynamics of Malaria in Humans and Mosquitoes: A Mathematical Model Analysis},
  author={Gebremichael, Shewakena Mersha and Mekonnen, Temesgen Tibebu},
  journal={IOSR Journal of Mathematics},
  volume={15},
  number={5},
  pages={46-59},
  year={2019},
  publisher={IOSR}
}

@article{kim2019effects,
  title={Effects of climate change on Plasmodium vivax malaria transmission dynamics: A mathematical modeling approach},
  author={Kim, Jung Eun and Choi, Yongin and Lee, Chang Hyeong},
  journal={Applied Mathematics and Computation},
  volume={347},
  pages={616--630},
  year={2019},
  publisher={Elsevier}
}

@article{silal2019malaria,
  title={Malaria elimination transmission and costing in the Asia-Pacific: a multi-species dynamic transmission model},
  author={Silal, Sheetal Prakash and Shretta, Rima and Celhay, Olivier J and Gran Mercado, Chris Erwin and Saralamba, Sompob and Maude, Richard James and White, Lisa Jane},
  journal={Wellcome Open Research},
  volume={4},
  pages={62},
  year={2019},
  publisher={F1000 Research Limited London, UK}
}

@article{kim2020mathematical,
  title={A mathematical model for assessing the effectiveness of controlling relapse in Plasmodium vivax malaria endemic in the Republic of Korea},
  author={Kim, Sungchan and Byun, Jong Hyuk and Park, Anna and Jung, Il Hyo},
  journal={PLoS One},
  volume={15},
  number={1},
  pages={e0227919},
  year={2020},
  publisher={Public Library of Science San Francisco, CA USA}
}

@article{corder2020modelling,
  title={Modelling the epidemiology of residual Plasmodium vivax malaria in a heterogeneous host population: a case study in the Amazon Basin},
  author={Corder, Rodrigo M and Ferreira, Marcelo U and Gomes, M Gabriela M},
  journal={PLoS Computational Biology},
  volume={16},
  number={3},
  pages={e1007377},
  year={2020},
  publisher={Public Library of Science San Francisco, CA USA}
}

@article{nyaberi2020mathematical,
  title={Mathematical Modeling of the Dynamics of Infectious Disease With Relapse},
  author={Nyaberi, Halson Ogeto and Wakwabubi, Vivtor Wangila},
  journal={Asian Journal of Mathematics and Computer Research},
  pages={28--37},
  year={2020}
}

@article{ghosh2020mathematical,
  title={Mathematical analysis of reinfection and relapse in malaria dynamics},
  author={Ghosh, Mini and Olaniyi, Samson and Obabiyi, Olawale S},
  journal={Applied Mathematics and Computation},
  volume={373},
  pages={125044},
  year={2020},
  publisher={Elsevier}
}

@inproceedings{baihaqi2020modelling,
  title={Modelling malaria transmission in a population with SEIRSp method},
  author={Baihaqi, Muhamad Adzib and Adi-Kusumo, Fajar},
  booktitle={AIP Conference Proceedings},
  volume={2264},
  number={1},
  year={2020},
  organization={AIP Publishing}
}

@article{ahkrizal2023dynamics,
  title={Dynamics System in the SEIR-SI Model of the Spread of Malaria with Recurrence},
  author={Ahkrizal, Afdhal and Jaharuddin, Jaharuddin and Nugrahani, Endar H},
  journal={Jambura Journal of Biomathematics (JJBM)},
  volume={4},
  number={1},
  pages={31--36},
  year={2023}
}

@article{aldila2021superinfection,
  title={A superinfection model on malaria transmission: analysis on the invasion basic reproduction number},
  author={Aldila, Dipo},
  journal={Commun. Math. Biol. Neurosci.},
  volume={2021},
  pages={Article--ID},
  year={2021}
}

@article{abimbade2022recurrent,
  title={Recurrent malaria dynamics: insight from mathematical modelling},
  author={Abimbade, Sulaimon F and Olaniyi, Samson and Ajala, Olusegun A},
  journal={The European Physical Journal Plus},
  volume={137},
  number={3},
  pages={292},
  year={2022},
  publisher={Springer}
}

@article{tasman2022assessing,
  title={Assessing the Impact of Relapse, Reinfection and Recrudescence on Malaria Eradication Policy: A Bifurcation and Optimal Control Analysis},
  author={Tasman, Hengki and Aldila, Dipo and Dumbela, Putri A and Ndii, Meksianis Z and Herdicho, Faishal F and Chukwu, Chidozie W},
  journal={Tropical Medicine and Infectious Disease},
  volume={7},
  number={10},
  pages={263},
  year={2022},
  publisher={MDPI}
}

@article{anwar2022multiscale,
  title={A Multiscale Mathematical Model of Plasmodium Vivax Transmission},
  author={Anwar, Md Nurul and Hickson, Roslyn I and Mehra, Somya and McCaw, James M and Flegg, Jennifer A},
  journal={Bulletin of Mathematical Biology},
  volume={84},
  number={8},
  pages={1--24},
  year={2022},
  publisher={Springer}
}

@article{mehra2022hypnozoite,
  title={Hypnozoite dynamics for Plasmodium vivax malaria: the epidemiological effects of radical cure},
  author={Mehra, Somya and Stadler, Eva and Khoury, David and McCaw, James M and Flegg, Jennifer A},
  journal={Journal of Theoretical Biology},
  pages={111014},
  year={2022},
  publisher={Elsevier}
}

@inproceedings{olaniyi2023optimal,
  title={Optimal Control Analysis of a Mathematical Model for Recurrent Malaria Dynamics},
  author={Olaniyi, Samson and Ajala, Olusegun A and Abimbade, Sulaimon F},
  booktitle={Operations Research Forum},
  volume={4},
  number={1},
  pages={14},
  year={2023},
  organization={Springer}
}

@article{anwar2023optimal,
   title={Optimal interruption of P. vivax malaria transmission using mass drug administration},
  author={Anwar, Md Nurul and Hickson, Roslyn I and Mehra, Somya and Price, David J and McCaw, James M and Flegg, Mark B and Flegg, Jennifer A},
  journal={Bulletin of Mathematical Biology},
  volume={85},
  number={6},
  pages={43},
  year={2023},
  publisher={Springer}
}

@article{olaniyi2023efficiency,
  title={Efficiency and economic analysis of intervention strategies for recurrent malaria transmission},
  author={Olaniyi, Samson and Abimbade, Sulaimon F and Ajala, Olusegun A and Chuma, Furaha M},
  journal={Quality \& Quantity},
  pages={1--19},
  year={2023},
  publisher={Springer}
}

@article{pizzitutti2015validated,
  title={A validated agent-based model to study the spatial and temporal heterogeneities of malaria incidence in the rainforest environment},
  author={Pizzitutti, Francesco and Pan, William and Barbieri, Alisson and Miranda, J Jaime and Feingold, Beth and Guedes, Gilvan R and Alarcon-Valenzuela, Javiera and Mena, Carlos F},
  journal={Malaria Journal},
  volume={14},
  pages={1--19},
  year={2015},
  publisher={Springer}
}

@article{white2018mathematical,
  title={Mathematical modelling of the impact of expanding levels of malaria control interventions on Plasmodium vivax},
  author={White, Michael T and Walker, Patrick and Karl, Stephan and Hetzel, Manuel W and Freeman, Tim and Waltmann, Andreea and Laman, Moses and Robinson, Leanne J and Ghani, Azra and Mueller, Ivo},
  journal={Nature Communications},
  volume={9},
  number={1},
  pages={1--10},
  year={2018},
  publisher={Nature Publishing Group}
}

@article{gharakhanlou2019developing,
  title={Developing an agent-based model for simulating the dynamic spread of Plasmodium vivax malaria: A case study of Sarbaz, Iran},
  author={Gharakhanlou, Navid Mahdizadeh and Mesgari, Mohammad Saadi and Hooshangi, Navid},
  journal={Ecological Informatics},
  volume={54},
  pages={101006},
  year={2019},
  publisher={Elsevier}
}

@article{nekkab2021estimated,
  title={Estimated impact of tafenoquine for Plasmodium vivax control and elimination in Brazil: A modelling study},
  author={Nekkab, Narimane and Lana, Raquel and Lacerda, Marcus and Obadia, Thomas and Siqueira, Andr{\'e} and Monteiro, Wuelton and Villela, Daniel and Mueller, Ivo and White, Michael},
  journal={PLoS Medicine},
  volume={18},
  number={4},
  pages={e1003535},
  year={2021},
  publisher={Public Library of Science San Francisco, CA USA}
}

@article{huber2021radical,
  title={How radical is radical cure? Site-specific biases in clinical trials underestimate the effect of radical cure on Plasmodium vivax hypnozoites},
  author={Huber, John H and Koepfli, Cristian and Espa{\~n}a, Guido and Nekkab, Narimane and White, Michael T and Alex Perkins, T},
  journal={Malaria Journal},
  volume={20},
  number={1},
  pages={1--15},
  year={2021},
  publisher={Springer}
}

@article{obadia2022developing,
  title={Developing sero-diagnostic tests to facilitate Plasmodium vivax Serological Test-and-Treat approaches: modeling the balance between public health impact and overtreatment},
  author={Obadia, Thomas and Nekkab, Narimane and Robinson, Leanne J and Drakeley, Chris and Mueller, Ivo and White, Michael T},
  journal={BMC Medicine},
  volume={20},
  number={1},
  pages={98},
  year={2022},
  publisher={Springer}
}

@article{white2022potential,
  title={Potential role of vaccines in elimination of Plasmodium vivax},
  author={White, Michael and Chitnis, Chetan E},
  journal={Parasitology International},
  volume={90},
  pages={102592},
  year={2022},
  publisher={Elsevier}
}

@article{walker2023model,
	title = {A model for malaria treatment evaluation in the presence of multiple species},
 	author = {Walker, C. R. and Hickson, R. I. and Chang, E. and Ngor, P. and Sovannaroth, S. and Simpson, J. A. and Price, D. J. and {McCaw}, J. M. and Price, R. N. and Flegg, J. A. and Devine, A.},
	journal = {Epidemics},
	volume = {44},
 	pages = {100687},
  year={2023},
  publisher={Elsevier}
}

@article{nekkab2023accelerating,
  title={Accelerating towards P. vivax elimination with a novel serological test-and-treat strategy: a modelling case study in Brazil},
  author={Nekkab, Narimane and Obadia, Thomas and Monteiro, Wuelton M and Lacerda, Marcus VG and White, Michael and Mueller, Ivo},
  journal={The Lancet Regional Health--Americas},
  volume={22},
  year={2023},
  publisher={Elsevier}
}

@article{kebaier2009kinetics,
  title={Kinetics of mosquito-injected Plasmodium sporozoites in mice: fewer sporozoites are injected into sporozoite-immunized mice},
  author={Kebaier, Chahnaz and Voza, Tatiana and Vanderberg, Jerome},
  journal={PLoS Pathogens},
  volume={5},
  number={4},
  year={2009},
  publisher={Public Library of Science}
}

@article{jones2006malaria,
  title={Malaria parasites up close},
  author={Jones, Malcolm K and Good, Michael F},
  journal={Nature Medicine},
  volume={12},
  number={2},
  pages={170--171},
  year={2006},
  publisher={Nature Publishing Group}
}

@article{price2007vivax,
  title={Vivax malaria: neglected and not benign},
  author={Price, Ric N and Tjitra, Emiliana and Guerra, Carlos A and Yeung, Shunmay and White, Nicholas J and Anstey, Nicholas M},
  journal={The American Journal of Tropical Medicine and Hygiene},
  volume={77},
  number={6\_Suppl},
  pages={79--87},
  year={2007},
  publisher={ASTMH}
}

@misc{baird2008real,
  title={Real-world therapies and the problem of vivax malaria},
  author={Baird, J Kevin},
  year={2008},
  publisher={Mass Medical Soc}
}

@article{betuela2012relapses,
  title={Relapses contribute significantly to the risk of Plasmodium vivax infection and disease in Papua New Guinean children 1--5 years of age},
  author={Betuela, Inoni and Rosanas-Urgell, Anna and Kiniboro, Benson and Stanisic, Danielle I and Samol, Lornah and de Lazzari, Elisa and del Portillo, Hernando A and Siba, Peter and Alonso, Pedro L and Bassat, Quique and others},
  journal={The Journal of Infectious Diseases},
  volume={206},
  number={11},
  pages={1771--1780},
  year={2012},
  publisher={Oxford University Press}
}

@article{luxemburger1999treatment,
  title={Treatment of vivax malaria on the western border of Thailand},
  author={Luxemburger, Christine and van Vugt, Mich{\`e}le and Jonathan, Saw and McGready, Rose and Looareesuwan, Sornchai and White, Nicholas J and Nosten, Fran{\c{c}}ois},
  journal={Transactions of the Royal Society of Tropical Medicine and Hygiene},
  volume={93},
  number={4},
  pages={433--438},
  year={1999},
  publisher={Royal Society of Tropical Medicine and Hygiene}
}

@article{breman2007defining,
  title={Defining and defeating the intolerable burden of malaria III. Progress and perspectives},
  author={Breman, Joel G and Alilio, Martin S and White, Nicholas J},
  journal={The American Journal of Tropical Medicine and Hygiene},
  volume={77},
  number={6\_Suppl},
  pages={vi--xi},
  year={2007},
  publisher={ASTMH}
}

@article{tjitra2008multidrug,
  title={Multidrug-resistant Plasmodium vivax associated with severe and fatal malaria: a prospective study in Papua, Indonesia},
  author={Tjitra, Emiliana and Anstey, Nicholas M and Sugiarto, Paulus and Warikar, Noah and Kenangalem, Enny and Karyana, Muhammad and Lampah, Daniel A and Price, Ric N},
  journal={PLoS Medicine},
  volume={5},
  number={6},
  year={2008},
  publisher={Public Library of Science}
}

@article{kochar2009severe,
  title={Severe Plasmodium vivax malaria: a report on serial cases from Bikaner in northwestern India},
  author={Kochar, Dhanpat K and Das, Ashish and Kochar, Sanjay K and Saxena, Vishal and Sirohi, Parmendra and Garg, Shilpi and Kochar, Abhishek and Khatri, Mahesh P and Gupta, Vikas},
  journal={The American Journal of Tropical Medicine and Hygiene},
  volume={80},
  number={2},
  pages={194--198},
  year={2009},
  publisher={ASTMH}
}

@book{keeling2011modeling,
  title={Modeling infectious diseases in humans and animals},
  author={Keeling, Matt J and Rohani, Pejman},
  year={2011},
  publisher={Princeton University Press}
}

@article{macdonald1957epidemiology,
  title={The epidemiology and control of malaria.},
  author={Macdonald, George and others},
  journal={The Epidemiology and Control of Malaria.},
  year={1957},
  publisher={Oxford University Press}
}

@article{smith2010quantitative,
  title={A quantitative analysis of transmission efficiency versus intensity for malaria},
  author={Smith, David L and Drakeley, Chris J and Chiyaka, Christinah and Hay, Simon I},
  journal={Nature Communications},
  volume={1},
  number={1},
  pages={1--9},
  year={2010},
  publisher={Nature Publishing Group}
}

@article{battle2014geographical,
  title={Geographical variation in Plasmodium vivax relapse},
  author={Battle, Katherine E and Karhunen, Markku S and Bhatt, Samir and Gething, Peter W and Howes, Rosalind E and Golding, Nick and Van Boeckel, Thomas P and Messina, Jane P and Shanks, G Dennis and Smith, David L and others},
  journal={Malaria Journal},
  volume={13},
  number={1},
  pages={144},
  year={2014},
  publisher={BioMed Central}
}

@article{portugal2011superinfection,
  title={Superinfection in malaria: Plasmodium shows its iron will},
  author={Portugal, S{\'\i}lvia and Drakesmith, Hal and Mota, Maria M},
  journal={EMBO Reports},
  volume={12},
  number={12},
  pages={1233--1242},
  year={2011},
  publisher={John Wiley \& Sons, Ltd}
}

@article{macdonald1950,
  title={},
  author={Macdonald, G.},
  journal={Tropical Diseases Bulletin},
  volume={47 (10)},
  pages={907-915},
  year={1950},
  publisher={}
}

@article{hulden2011activation,
  title={Activation of the hypnozoite: a part of Plasmodium vivax life cycle and survival},
  author={Hulden, Lena and Hulden, Larry},
  journal={Malaria Journal},
  volume={10},
  number={1},
  pages={90},
  year={2011},
  publisher={Springer}
}

@article{watson2018implications,
  title={Implications of current therapeutic restrictions for primaquine and tafenoquine in the radical cure of vivax malaria},
  author={Watson, James and Taylor, Walter RJ and Bancone, Germana and Chu, Cindy S and Jittamala, Podjanee and White, Nicholas J},
  journal={PLoS Neglected Tropical Diseases},
  volume={12},
  number={4},
  pages={e0006440},
  year={2018},
  publisher={Public Library of Science}
}

@article{mehra2020activation,
  title={An Activation-Clearance Model for Plasmodium vivax Malaria},
  author={Mehra, Somya and McCaw, James M and Flegg, Mark B and Taylor, Peter G and Flegg, Jennifer A},
  journal={Bulletin of Mathematical Biology},
  volume={82},
  number={2},
  pages={32},
  year={2020},
  publisher={Springer}
}

@article{adekunle2015modeling,
  title={Modeling the dynamics of Plasmodium vivax infection and hypnozoite reactivation in vivo},
  author={Adekunle, Adeshina I and Pinkevych, Mykola and McGready, Rose and Luxemburger, Christine and White, Lisa J and Nosten, Francois and Cromer, Deborah and Davenport, Miles P},
  journal={PLoS Neglected Tropical Diseases},
  volume={9},
  number={3},
  year={2015},
  publisher={Public Library of Science}
}

@article{howes2012g6pd,
  title={G6PD deficiency prevalence and estimates of affected populations in malaria endemic countries: a geostatistical model-based map},
  author={Howes, Rosalind E and Piel, Frederic B and Patil, Anand P and Nyangiri, Oscar A and Gething, Peter W and Dewi, Mewahyu and Hogg, Mariana M and Battle, Katherine E and Padilla, Carmencita D and Baird, J Kevin and others},
  journal={PLoS Medicine},
  volume={9},
  number={11},
  year={2012},
  publisher={Public Library of Science}
}

@article{white2011determinants,
  title={Determinants of relapse periodicity in Plasmodium vivax malaria},
  author={White, Nicholas J},
  journal={Malaria Journal},
  volume={10},
  number={1},
  pages={297},
  year={2011},
  publisher={Springer}
}

@article{doolan2009acquired,
  title={Acquired immunity to malaria},
  author={Doolan, Denise L and Doba{\~n}o, Carlota and Baird, J Kevin},
  journal={Clinical Microbiology Reviews},
  volume={22},
  number={1},
  pages={13--36},
  year={2009},
  publisher={Am Soc Microbiol}
}

@article{antinori2012biology,
  title={Biology of human malaria plasmodia including Plasmodium knowlesi},
  author={Antinori, Spinello and Galimberti, Laura and Milazzo, Laura and Corbellino, Mario},
  journal={Mediterranean Journal of Hematology and Infectious Diseases},
  volume={4},
  number={1},
  year={2012},
  publisher={Catholic University in Rome}
}

@article{world2020world,
  title={World malaria report 2020: 20 years of global progress and challenges},
  author={WHO},
  year={2020},
  publisher={World Health Organization}
}

@article{imwong2007relapses,
  title={Relapses of Plasmodium vivax infection usually result from activation of heterologous hypnozoites},
  author={Imwong, Mallika and Snounou, Georges and Pukrittayakamee, Sasithon and Tanomsing, Naowarat and Kim, Jung Ryong and Nandy, Amitab and Guthmann, Jean-Paul and Nosten, Francois and Carlton, Jane and Looareesuwan, Sornchai and others},
  journal={The Journal of Infectious Diseases},
  volume={195},
  number={7},
  pages={927--933},
  year={2007},
  publisher={The University of Chicago Press}
}

@article{huppert2013mathematical,
  title={Mathematical modelling and prediction in infectious disease epidemiology},
  author={Huppert, Amit and Katriel, Guy},
  journal={Clinical Microbiology and Infection},
  volume={19},
  number={11},
  pages={999--1005},
  year={2013},
  publisher={Elsevier}
}

@article{barnabas2006epidemiology,
  title={Epidemiology of HPV 16 and cervical cancer in Finland and the potential impact of vaccination: mathematical modelling analyses},
  author={Barnabas, Ruanne V and Laukkanen, P{\"a}ivi and Koskela, Pentti and Kontula, Osmo and Lehtinen, Matti and Garnett, Geoff P},
  journal={PLoS Med},
  volume={3},
  number={5},
  pages={e138},
  year={2006},
  publisher={Public Library of Science}
}

@article{commons2019risk,
  title={Risk of Plasmodium vivax parasitaemia after Plasmodium falciparum infection: a systematic review and meta-analysis},
  author={Commons, Robert J and Simpson, Julie A and Thriemer, Kamala and Hossain, Mohammad S and Douglas, Nicholas M and Humphreys, Georgina S and Sibley, Carol H and Guerin, Philippe J and Price, Ric N},
  journal={The Lancet Infectious Diseases},
  volume={19},
  number={1},
  pages={91--101},
  year={2019},
  publisher={Elsevier}
}

@article{commons2018effect,
  title={The effect of chloroquine dose and primaquine on Plasmodium vivax recurrence: a WorldWide Antimalarial Resistance Network systematic review and individual patient pooled meta-analysis},
  author={Commons, Robert J and Simpson, Julie A and Thriemer, Kamala and Humphreys, Georgina S and Abreha, Tesfay and Alemu, Sisay G and A{\~n}ez, Arletta and Anstey, Nicholas M and Awab, Ghulam R and Baird, J Kevin and others},
  journal={The Lancet Infectious Diseases},
  volume={18},
  number={9},
  pages={1025--1034},
  year={2018},
  publisher={Elsevier}
}

@article{mehra2021antibody,
  title={Antibody Dynamics for Plasmodium vivax Malaria: A Mathematical Model},
  author={Mehra, Somya and McCaw, James M and Flegg, Mark B and Taylor, Peter G and Flegg, Jennifer A},
  journal={Bulletin of Mathematical Biology},
  volume={83},
  number={1},
  pages={1--27},
  year={2021},
  publisher={Springer}
}

@article{wells2010targeting,
  title={Targeting the hypnozoite reservoir of Plasmodium vivax: the hidden obstacle to malaria elimination},
  author={Wells, Timothy NC and Burrows, Jeremy N and Baird, J Kevin},
  journal={Trends in Parasitology},
  volume={26},
  number={3},
  pages={145--151},
  year={2010},
  publisher={Elsevier}
}

@article{battle2019mapping,
  title={Mapping the global endemicity and clinical burden of Plasmodium vivax, 2000--17: a spatial and temporal modelling study},
  author={Battle, Katherine E and Lucas, Tim CD and Nguyen, Michele and Howes, Rosalind E and Nandi, Anita K and Twohig, Katherine A and Pfeffer, Daniel A and Cameron, Ewan and Rao, Puja C and Casey, Daniel and others},
  journal={The Lancet},
  volume={394},
  number={10195},
  pages={332--343},
  year={2019},
  publisher={Elsevier}
}

@article{kim2012genotyping,
  title={Genotyping of Plasmodium vivax reveals both short and long latency relapse patterns in Kolkata},
  author={Kim, Jung-Ryong and Nandy, Amitabha and Maji, Ardhendu Kumar and Addy, Manjulika and Dondorp, Arjen M and Day, Nicholas PJ and Pukrittayakamee, Sasithon and White, Nicholas J and Imwong, Mallika},
  journal={PloS One},
  volume={7},
  number={7},
  pages={e39645},
  year={2012},
  publisher={Public Library of Science San Francisco, USA}
}

@article{newby2015review,
  title={Review of mass drug administration for malaria and its operational challenges},
  author={Newby, Gretchen and Hwang, Jimee and Koita, Kadiatou and Chen, Ingrid and Greenwood, Brian and Von Seidlein, Lorenz and Shanks, G Dennis and Slutsker, Laurence and Kachur, S Patrick and Wegbreit, Jennifer and others},
  journal={The American Journal of Tropical Medicine and Hygiene},
  volume={93},
  number={1},
  pages={125},
  year={2015},
  publisher={The American Society of Tropical Medicine and Hygiene}
}

@article{price2020plasmodium,
  title={Plasmodium vivax in the Era of the Shrinking P. falciparum Map},
  author={Price, Ric N and Commons, Robert J and Battle, Katherine E and Thriemer, Kamala and Mendis, Kamini},
  journal={Trends in Parasitology},
  volume={36},
  number={6},
  pages={560--570},
  year={2020},
  publisher={Elsevier}
}

@article{hsiang2013mass,
  title={Mass drug administration for the control and elimination of Plasmodium vivax malaria: an ecological study from Jiangsu province, China},
  author={Hsiang, Michelle S and Hwang, Jimee and Tao, Amy R and Liu, Yaobao and Bennett, Adam and Shanks, George Dennis and Cao, Jun and Kachur, Stephen Patrick and Feachem, Richard GA and Gosling, Roly D and others},
  journal={Malaria Journal},
  volume={12},
  number={1},
  pages={1--14},
  year={2013},
  publisher={BioMed Central}
}

@article{popovici2015challenges,
  title={Challenges in antimalarial drug treatment for vivax malaria control},
  author={Popovici, Jean and M{\'e}nard, Didier},
  journal={Trends in Molecular Medicine},
  volume={21},
  number={12},
  pages={776--788},
  year={2015},
  publisher={Elsevier}
}

@book{world2022world,
  title={World malaria report 2022},
  author={WHO},
  year={2022},
  publisher={World Health Organization}
}

@article{devine2021global,
  title={Global economic costs due to vivax malaria and the potential impact of its radical cure: A modelling study},
  author={Devine, Angela and Battle, Katherine E and Meagher, Niamh and Howes, Rosalind E and Dini, Saber and Gething, Peter W and Simpson, Julie A and Price, Ric N and Lubell, Yoel},
  journal={PLoS Medicine},
  volume={18},
  number={6},
  pages={e1003614},
  year={2021},
  publisher={Public Library of Science San Francisco, CA USA}
}

@article{baird2013evidence,
  title={Evidence and implications of mortality associated with acute Plasmodium vivax malaria},
  author={Baird, J Kevin},
  journal={Clinical Microbiology Reviews},
  volume={26},
  number={1},
  pages={36--57},
  year={2013},
  publisher={Am Soc Microbiol}
}

@article{zuber2018multidrug,
  title={Multidrug-resistant malaria and the impact of mass drug administration},
  author={Zuber, Janie Anne and Takala-Harrison, Shannon},
  journal={Infection and Drug Resistance},
  volume={11},
  pages={299},
  year={2018},
  publisher={Dove Press}
}

@article{commons2020estimating,
  title={Estimating the proportion of Plasmodium vivax recurrences caused by relapse: a systematic review and meta-analysis},
  author={Commons, Robert J and Simpson, Julie A and Watson, James and White, Nicholas J and Price, Ric N},
  journal={The American Journal of Tropical Medicine and Hygiene},
  volume={103},
  number={3},
  pages={1094},
  year={2020},
  publisher={The American Society of Tropical Medicine and Hygiene}
}

@article{taylor2019short,
  title={Short-course primaquine for the radical cure of Plasmodium vivax malaria: a multicentre, randomised, placebo-controlled non-inferiority trial},
  author={Taylor, Walter RJ and Thriemer, Kamala and von Seidlein, Lorenz and Yuentrakul, Prayoon and Assawariyathipat, Thanawat and Assefa, Ashenafi and Auburn, Sarah and Chand, Krisin and Chau, Nguyen Hoang and Cheah, Phaik Yeong and others},
  journal={The Lancet},
  volume={394},
  number={10202},
  pages={929--938},
  year={2019},
  publisher={Elsevier}
}

@article{poespoprodjo2022supervised,
  title={Supervised versus unsupervised primaquine radical cure for the treatment of falciparum and vivax malaria in Papua, Indonesia: a cluster-randomised, controlled, open-label superiority trial},
  author={Poespoprodjo, Jeanne Rini and Burdam, Faustina Helena and Candrawati, Freis and Ley, Benedikt and Meagher, Niamh and Kenangalem, Enny and Indrawanti, Ratni and Trianty, Leily and Thriemer, Kamala and Price, David J and others},
  journal={The Lancet Infectious Diseases},
  volume={22},
  number={3},
  pages={367--376},
  year={2022},
  publisher={Elsevier}
}

@article{world2021second,
  title={Second focused review meeting of the Malaria Elimination Oversight Committee (MEOC): report of a virtual meeting, 28 June--1 July 2021},
  author={WHO},
  year={2021},
  publisher={World Health Organization}
}

@book{beran1994statistics,
  title={Statistics for long-memory processes},
  author={Beran, Jan},
  volume={61},
  year={1994},
  publisher={CRC press}
}

@article{allen2000comparison,
  title={Comparison of deterministic and stochastic SIS and SIR models in discrete time},
  author={Allen, Linda JS and Burgin, Amy M},
  journal={Mathematical Biosciences},
  volume={163},
  number={1},
  pages={1--33},
  year={2000},
  publisher={Elsevier}
}

@misc{volker2005individual,
  title={Individual-based modeling and ecology},
  author={Volker, Grimm and Railsback, Steven F},
  year={2005},
  publisher={Princeton University Press, Princeton}
}

@article{crooks2008key,
  title={Key challenges in agent-based modelling for geo-spatial simulation},
  author={Crooks, Andrew and Castle, Christian and Batty, Michael},
  journal={Computers, Environment and Urban Systems},
  volume={32},
  number={6},
  pages={417--430},
  year={2008},
  publisher={Elsevier}
}

@misc{colman1998complexity,
  title={The complexity of cooperation: Agent-based models of competition and collaboration},
  author={Colman, Andrew M},
  year={1998},
  publisher={John Wiley \& Sons, Inc. New York}
}

@article{samuelson2006agent,
  title={Agent-Based Simulation Comes of Age: Software opens up many new areas of application},
  author={Samuelson, Douglas A and Macal, Charles M},
  journal={OR/MS Today},
  volume={33},
  number={4},
  pages={34--39},
  year={2006},
  publisher={Institute for Operations Research and the Management Sciences}
}

@article{bonabeau2002agent,
  title={Agent-based modeling: Methods and techniques for simulating human systems},
  author={Bonabeau, Eric},
  journal={Proceedings of The National Academy of Sciences},
  volume={99},
  number={suppl\_3},
  pages={7280--7287},
  year={2002},
  publisher={National Acad Sciences}
}

@article{laubenbacher2020agent,
  title={Agent-based modeling, Mathematical formalism for},
  author={Laubenbacher, Reinhard and Jarrah, Abdul S and Mortveit, Henning S and Ravi, SS},
  journal={Complex Social and Behavioral Systems: Game Theory and Agent-Based Models},
  pages={683--703},
  year={2020},
  publisher={Springer}
}

@article{mohd2022revisiting,
  title={Revisiting discrepancies between stochastic agent-based and deterministic models},
  author={Mohd, Mohd Hafiz},
  journal={Community Ecology},
  volume={23},
  number={3},
  pages={453--468},
  year={2022},
  publisher={Springer}
}

@misc{garabed2020multi,
  title={Multi-scale dynamics of infectious diseases},
  author={Garabed, Rebecca B and Jolles, Anna and Garira, Winston and Lanzas, Cristina and Gutierrez, Juan and Rempala, Grzegorz},
  journal={Interface Focus},
  volume={10},
  number={1},
  pages={20190118},
  year={2020},
  publisher={The Royal Society}
}

@article{shanks2013activation,
  title={The activation of vivax malaria hypnozoites by infectious diseases},
  author={Shanks, G Dennis and White, Nicholas J},
  journal={The Lancet Infectious Diseases},
  volume={13},
  number={10},
  pages={900--906},
  year={2013},
  publisher={Elsevier}
}

@article{chhim2021malaria,
  title={Malaria in Cambodia: a retrospective analysis of a changing epidemiology 2006--2019},
  author={Chhim, Srean and Piola, Patrice and Housen, Tambri and Herbreteau, Vincent and Tol, Bunkea},
  journal={International Journal of Environmental Research and Public Health},
  volume={18},
  number={4},
  pages={1960},
  year={2021},
  publisher={MDPI}
}

@article{snounou2004co,
  title={The co-existence of Plasmodium: sidelights from falciparum and vivax malaria in Thailand},
  author={Snounou, Georges and White, Nicolas J},
  journal={Trends in Parasitology},
  volume={20},
  number={7},
  pages={333--339},
  year={2004},
  publisher={Elsevier}
}

@article{lin2011plasmodium,
  title={Plasmodium falciparum gametocyte carriage is associated with subsequent Plasmodium vivax relapse after treatment},
  author={Lin, Jessica T and Bethell, Delia and Tyner, Stuart D and Lon, Chanthap and Shah, Naman K and Saunders, David L and Sriwichai, Sabaithip and Khemawoot, Phisit and Kuntawunggin, Worachet and Smith, Bryan L and others},
  journal={PLoS One},
  volume={6},
  number={4},
  pages={e18716},
  year={2011},
  publisher={Public Library of Science San Francisco, USA}
}

@article{langhorne2008immunity,
  title={Immunity to malaria: more questions than answers},
  author={Langhorne, Jean and Ndungu, Francis M and Sponaas, Anne-Marit and Marsh, Kevin},
  journal={Nature Immunology},
  volume={9},
  number={7},
  pages={725--732},
  year={2008},
  publisher={Nature Publishing Group US New York}
}

@article{port1980relationship,
  title={The relationship of host size to feeding by mosquitoes of the Anopheles gambiae Giles complex (Diptera: Culicidae)},
  author={Port, GR and Boreham, PFL and Bryan, Joan H},
  journal={Bulletin of Entomological Research},
  volume={70},
  number={1},
  pages={133--144},
  year={1980},
  publisher={Cambridge University Press}
}

@article{carnevale1978aggressiveness,
  title={The aggressiveness of Anopheles gambiae A in relation to the age and sex of the human subjects},
  author={Carnevale, P and Frezil, JL and Bosseno, MF and Le Pont, F and Lancien, J},
  journal={Bulletin of the World Health Organization},
  volume={56},
  number={1},
  pages={147--154},
  year={1978}
}

@article{singh2022mass,
  title={Mass screening and treatment (MSaT) for identifying and treating asymptomatic cases of malaria-malaria elimination demonstration project (MEDP), Mandla, Madhya Pradesh},
  author={Singh, Akansha and Rajvanshi, Harsh and Singh, Mrigendra P and Bhandari, Sneha and Nisar, Sekh and Poriya, Rajan and Telasey, Vinay and Jayswar, Himanshu and Mishra, Ashok K and Das, Aparup and others},
  journal={Malaria Journal},
  volume={21},
  number={1},
  pages={1--9},
  year={2022},
  publisher={BioMed Central}
}

@article{kim2021systematic,
  title={A systematic review of the evidence on the effectiveness and cost-effectiveness of mass screen-and-treat interventions for malaria control},
  author={Kim, Sooyoung and Luande, Verah Nafula and Rockl{\"o}v, Joacim and Carlton, Jane M and Tozan, Yesim},
  journal={The American Journal of Tropical Medicine and Hygiene},
  volume={105},
  number={6},
  pages={1722},
  year={2021},
  publisher={The American Society of Tropical Medicine and Hygiene}
}

@article{smith2012ross,
  title={Ross, Macdonald, and a theory for the dynamics and control of mosquito-transmitted pathogens},
  author={Smith, David L and Battle, Katherine E and Hay, Simon I and Barker, Christopher M and Scott, Thomas W and McKenzie, F Ellis},
  journal={PLoS Pathogens},
  volume={8},
  number={4},
  pages={e1002588},
  year={2012},
  publisher={Public Library of Science San Francisco, USA}
}

@MASTERSTHESIS{thesis_somya,
  AUTHOR =       {Somya Mehra},
  TYPE     = {MSc thesis},
  TITLE =        {Epidemic models for malaria:
superinfection},
  SCHOOL =       {The University of Melbourne},
  YEAR =         {2022},
  month =        {},
  }

@article{bashar2014seasonal,
  title={Seasonal abundance of Anopheles mosquitoes and their association with meteorological factors and malaria incidence in Bangladesh},
  author={Bashar, Kabirul and Tuno, Nobuko},
  journal={Parasites \& Vectors},
  volume={7},
  number={1},
  pages={1--10},
  year={2014},
  publisher={BioMed Central}
}

@book{nasell2013hybrid,
  title={Hybrid models of tropical infections},
  author={Nasell, Ingemar},
  volume={59},
  year={2013},
  publisher={Springer Science \& Business Media}
}

@article{mayer2010microbiology,
  title={Microbiology and Immunology On-Line Textbook},
  author={Mayer, Gene},
  journal={USC School of Medicine. Available [online]< http://pathmicro. med. sc. edu/ghaffar/innate. htm>, retrieved on 10th May},
  year={2010}
}

@misc{chaffey2003alberts,
  title={Alberts, B., Johnson, A., Lewis, J., Raff, M., Roberts, K. and Walter, P. Molecular biology of the cell. 4th edn.},
  author={Chaffey, Nigel},
  year={2003},
  publisher={Oxford University Press}
}

@article{barua2019impact,
  title={The impact of early life exposure to Plasmodium falciparum on the development of naturally acquired immunity to malaria in young Malawian children},
  author={Barua, Priyanka and Beeson, James G and Maleta, Kenneth and Ashorn, Per and Rogerson, Stephen J},
  journal={Malaria Journal},
  volume={18},
  pages={1--12},
  year={2019},
  publisher={Springer}
}

@misc{Darrin_Darrin_2021, 
title={Pros \& Cons of Agent-based modeling}, url={https://educationalresearchtechniques.com/2020/08/07/pros-cons-of-agent-based-modeling/}, journal={Educational Research Techniques}, author={Darrin, Dr. and Darrin, Dr.}, year={2021}}

@article{recht2018use,
  title={Use of primaquine and glucose-6-phosphate dehydrogenase deficiency testing: divergent policies and practices in malaria endemic countries},
  author={Recht, Judith and Ashley, Elizabeth A and White, Nicholas J},
  journal={PLoS Neglected Tropical Diseases},
  volume={12},
  number={4},
  pages={e0006230},
  year={2018},
  publisher={Public Library of Science San Francisco, CA USA}
}

@book{bruce1985essential,
  title={Essential Malariology},
  author={Bruce-Chwatt, Leonard Jan and others},
  year={1985},
  publisher={William Heinemann Medical Books Ltd}
}

@article{champagne2022using,
  title={Using observed incidence to calibrate the transmission level of a mathematical model for Plasmodium vivax dynamics including case management and importation},
  author={Champagne, Clara and Gerhards, Maximilian and Lana, Justin and Espinosa, Bernardo Garc{\'\i}a and Bradley, Christina and Gonz{\'a}lez, Oscar and Cohen, Justin M and Le Menach, Arnaud and White, Michael T and Pothin, Emilie},
  journal={Mathematical Biosciences},
  volume={343},
  pages={108750},
  year={2022},
  publisher={Elsevier}
}

@article{mandal2011mathematical,
  title={Mathematical models of malaria-a review},
  author={Mandal, Sandip and Sarkar, Ram Rup and Sinha, Somdatta},
  journal={Malaria Journal},
  volume={10},
  number={1},
  pages={1--19},
  year={2011},
  publisher={BioMed Central}
}

@article{hossain_risk_2020,
	title = {The risk of Plasmodium vivax parasitaemia after P. falciparum malaria: An individual patient data meta-analysis from the {WorldWide} Antimalarial Resistance Network},
	volume = {17},
	issn = {1549-1676},
	doi = {10.1371/journal.pmed.1003393},
	pages = {e1003393},
	number = {11},
	journaltitle = {{PLOS} Medicine},
	author = {Hossain, Mohammad S. and Commons, Robert J. and Douglas, Nicholas M. and Thriemer, Kamala and Alemayehu, Bereket H. and Amaratunga, Chanaki and Anvikar, Anupkumar R. and Ashley, Elizabeth A. and Asih, Puji B. S. and Carrara, Verena I. and Lon, Chanthap and D’Alessandro, Umberto and Davis, Timothy M. E. and Dondorp, Arjen M. and Edstein, Michael D. and Fairhurst, Rick M. and Ferreira, Marcelo U. and Hwang, Jimee and Janssens, Bart and Karunajeewa, Harin and Kiechel, Jean R. and Ladeia-Andrade, Simone and Laman, Moses and Mayxay, Mayfong and {McGready}, Rose and Moore, Brioni R. and Mueller, Ivo and Newton, Paul N. and Thuy-Nhien, Nguyen T. and Noedl, Harald and Nosten, Francois and Phyo, Aung P. and Poespoprodjo, Jeanne R. and Saunders, David L. and Smithuis, Frank and Spring, Michele D. and Stepniewska, Kasia and Suon, Seila and Suputtamongkol, Yupin and Syafruddin, Din and Tran, Hien T. and Valecha, Neena and Herp, Michel Van and Vugt, Michele Van and White, Nicholas J. and Guerin, Philippe J. and Simpson, Julie A. and Price, Ric N.},
	date = {2020-11-19},
	langid = {english},
	note = {Publisher: Public Library of Science},
}

@article{commons_risk_2019,
	title = {Risk of Plasmodium vivax parasitaemia after Plasmodium falciparum infection: a systematic review and meta-analysis},
	volume = {19},
	issn = {1473-3099, 1474-4457},
	doi = {10.1016/S1473-3099(18)30596-6},
	pages = {91--101},
	number = {1},
	journaltitle = {The Lancet Infectious Diseases},
	author = {Commons, Robert J. and Simpson, Julie A. and Thriemer, Kamala and Hossain, Mohammad S. and Douglas, Nicholas M. and Humphreys, Georgina S. and Sibley, Carol H. and Guerin, Philippe J. and Price, Ric N.},
	date = {2019-01-01},
	pmid = {30587297},
	note = {Publisher: Elsevier},
}

@article{taylor_resolving_2019,
	title = {Resolving the cause of recurrent Plasmodium vivax malaria probabilistically},
	volume = {10},
	rights = {2019 The Author(s)},
	issn = {2041-1723},
	doi = {10.1038/s41467-019-13412-x},
	pages = {5595},
	number = {1},
	journaltitle = {Nature Communications},
	author = {Taylor, Aimee R. and Watson, James A. and Chu, Cindy S. and Puaprasert, Kanokpich and Duanguppama, Jureeporn and Day, Nicholas P. J. and Nosten, Francois and Neafsey, Daniel E. and Buckee, Caroline O. and Imwong, Mallika and White, Nicholas J.},
	date = {2019-12-06},
	langid = {english},
	note = {Number: 1
Publisher: Nature Publishing Group},
}

@article{lover2013quantifying,
  title={Quantifying effect of geographic location on epidemiology of Plasmodium vivax malaria},
  author={Lover, Andrew A and Coker, Richard J},
  journal={Emerging Infectious Diseases},
  volume={19},
  number={7},
  pages={1058},
  year={2013},
  publisher={Centers for Disease Control and Prevention}
}

@article{yao2015law,
  title={Law of large numbers for uncertain random variables},
  author={Yao, Kai and Gao, Jinwu},
  journal={IEEE Transactions on Fuzzy Systems},
  volume={24},
  number={3},
  pages={615--621},
  year={2015},
  publisher={IEEE}
}

@article{battle2021global,
  title={The global burden of Plasmodium vivax malaria is obscure and insidious},
  author={Battle, Katherine E and Baird, J Kevin},
  journal={PLoS Medicine},
  volume={18},
  number={10},
  pages={e1003799},
  year={2021},
  publisher={Public Library of Science}
}

@article{okell2011potential,
  title={The potential contribution of mass treatment to the control of Plasmodium falciparum malaria},
  author={Okell, Lucy C and Griffin, Jamie T and Kleinschmidt, Immo and Hollingsworth, T D{\'e}irdre and Churcher, Thomas S and White, Michael J and Bousema, Teun and Drakeley, Chris J and Ghani, Azra C},
  journal={PLOS One},
  volume={6},
  number={5},
  pages={e20179},
  year={2011},
  publisher={Public Library of Science San Francisco, USA}
}

@article{grueninger2013transitioning,
  title={Transitioning from malaria control to elimination: the vital role of ACTs},
  author={Grueninger, Heiner and Hamed, Kamal},
  journal={Trends in Parasitology},
  volume={29},
  number={2},
  pages={60--64},
  year={2013},
  publisher={Elsevier}
}

@article{keven2022vector,
  title={Vector composition, abundance, biting patterns and malaria transmission intensity in Madang, Papua New Guinea: assessment after 7 years of an LLIN-based malaria control programme},
  author={Keven, John B and Katusele, Michelle and Vinit, Rebecca and Rodr{\'\i}guez-Rodr{\'\i}guez, Daniela and Hetzel, Manuel W and Robinson, Leanne J and Laman, Moses and Karl, Stephan and Walker, Edward D},
  journal={Malaria Journal},
  volume={21},
  number={1},
  pages={1--15},
  year={2022},
  publisher={BioMed Central}
}

@article{reimer2016malaria,
  title={Malaria transmission dynamics surrounding the first nationwide long-lasting insecticidal net distribution in Papua New Guinea},
  author={Reimer, Lisa J and Thomsen, Edward K and Koimbu, Gussy and Keven, John B and Mueller, Ivo and Siba, Peter M and Kazura, James W and Hetzel, Manuel W and Zimmerman, Peter A},
  journal={Malaria Journal},
  volume={15},
  pages={1--11},
  year={2016},
  publisher={Springer}
}

@article{smith2018agent,
  title={Agent-based models of malaria transmission: a systematic review},
  author={Smith, Neal R and Trauer, James M and Gambhir, Manoj and Richards, Jack S and Maude, Richard J and Keith, Jonathan M and Flegg, Jennifer A},
  journal={Malaria Journal},
  volume={17},
  number={1},
  pages={1--16},
  year={2018},
  publisher={BioMed Central}
}

@article{stadler2022population,
  title={Population heterogeneity in Plasmodium vivax relapse risk},
  author={Stadler, Eva and Cromer, Deborah and Mehra, Somya and Adekunle, Adeshina I and Flegg, Jennifer A and Anstey, Nicholas M and Watson, James A and Chu, Cindy S and Mueller, Ivo and Robinson, Leanne J and others},
  journal={PLOS Neglected Tropical Diseases},
  volume={16},
  number={12},
  pages={e0010990},
  year={2022},
  publisher={Public Library of Science San Francisco, CA USA}
}

@article{joyner2019humoral,
  title={Humoral immunity prevents clinical malaria during Plasmodium relapses without eliminating gametocytes},
  author={Joyner, Chester J and Brito, Cristiana FA and Saney, Celia L and Joice Cordy, Regina and Smith, Maren L and Lapp, Stacey A and Cabrera-Mora, Monica and Kyu, Shuya and Lackman, Nicolas and Nural, Mustafa V and others},
  journal={PLoS Pathogens},
  volume={15},
  number={9},
  pages={e1007974},
  year={2019},
  publisher={Public Library of Science San Francisco, CA USA}
}

@article{hofmann2017complex,
  title={The complex relationship of exposure to new Plasmodium infections and incidence of clinical malaria in Papua New Guinea},
  author={Hofmann, Natalie E and Karl, Stephan and Wampfler, Rahel and Kiniboro, Benson and Teliki, Albina and Iga, Jonah and Waltmann, Andreea and Betuela, Inoni and Felger, Ingrid and Robinson, Leanne J and others},
  journal={Elife},
  volume={6},
  pages={e23708},
  year={2017},
  publisher={eLife Sciences Publications, Ltd}
}

@article{haque2010spatial,
  title={Spatial prediction of malaria prevalence in an endemic area of Bangladesh},
  author={Haque, Ubydul and Magalh{\~a}es, Ricardo J Soares and Reid, Heidi L and Clements, Archie CA and Ahmed, Syed Masud and Islam, Akramul and Yamamoto, Taro and Haque, Rashidul and Glass, Gregory E},
  journal={Malaria Journal},
  volume={9},
  number={1},
  pages={1--10},
  year={2010},
  publisher={BioMed Central}
}

@article{amratia2019characterizing,
  title={Characterizing local-scale heterogeneity of malaria risk: a case study in Bunkpurugu-Yunyoo district in northern Ghana},
  author={Amratia, Punam and Psychas, Paul and Abuaku, Benjamin and Ahorlu, Collins and Millar, Justin and Oppong, Samuel and Koram, Kwadwo and Valle, Denis},
  journal={Malaria Journal},
  volume={18},
  pages={1--14},
  year={2019},
  publisher={Springer}
}

@article{muh2020cross,
  title={Cross-species reactivity of antibodies against Plasmodium vivax blood-stage antigens to Plasmodium knowlesi},
  author={Muh, Fauzi and Kim, Namhyeok and Nyunt, Myat Htut and Firdaus, Egy Rahman and Han, Jin-Hee and Hoque, Mohammad Rafiul and Lee, Seong-Kyun and Park, Ji-Hoon and Moon, Robert W and Lau, Yee Ling and others},
  journal={PLoS Neglected Tropical Diseases},
  volume={14},
  number={6},
  pages={e0008323},
  year={2020},
  publisher={Public Library of Science San Francisco, CA USA}
}

@article{cox2008knowlesi,
  title={Knowlesi malaria: newly emergent and of public health importance?},
  author={Cox-Singh, Janet and Singh, Balbir},
  journal={Trends in Parasitology},
  volume={24},
  number={9},
  pages={406--410},
  year={2008},
  publisher={Elsevier}
}

@article{kotepui2020plasmodium,
  title={Plasmodium spp. mixed infection leading to severe malaria: A systematic review and meta-analysis},
  author={Kotepui, Manas and Kotepui, Kwuntida Uthaisar and De Jesus Milanez, Giovanni and Masangkay, Frederick Ramirez},
  journal={Scientific Reports},
  volume={10},
  number={1},
  pages={11068},
  year={2020},
  publisher={Nature Publishing Group UK London}
}

@article{mohapatra2012profile,
  title={Profile of mixed species (Plasmodium vivax and falciparum) malaria in adults},
  author={Mohapatra, MK and Dash, LK and Bariha, PK and Karua, PC},
  journal={J Assoc Physicians India},
  volume={60},
  pages={20--24},
  year={2012}
}

@article{stresman2020association,
  title={Association between the proportion of Plasmodium falciparum and Plasmodium vivax infections detected by passive surveillance and the magnitude of the asymptomatic reservoir in the community: a pooled analysis of paired health facility and community data},
  author={Stresman, Gillian and Sep{\'u}lveda, Nuno and Fornace, Kimberly and Grignard, Lynn and Mwesigwa, Julia and Achan, Jane and Miller, John and Bridges, Daniel J and Eisele, Thomas P and Mosha, Jacklin and others},
  journal={The Lancet Infectious Diseases},
  volume={20},
  number={8},
  pages={953--963},
  year={2020},
  publisher={Elsevier}
}

@article{mueller2013natural,
  title={Natural acquisition of immunity to Plasmodium vivax: epidemiological observations and potential targets},
  author={Mueller, Ivo and Galinski, Mary R and Tsuboi, Takafumi and Arevalo-Herrera, Myriam and Collins, William E and King, Christopher L},
  journal={Advances in Parasitology},
  volume={81},
  pages={77--131},
  year={2013},
  publisher={Elsevier}
}

@article{gruner2007sterile,
  title={Sterile protection against malaria is independent of immune responses to the circumsporozoite protein},
  author={Gr{\"u}ner, Anne Charlotte and Mauduit, Marjorie and Tewari, Rita and Romero, Jackeline F and Depinay, Nadya and Kayibanda, Mich{\`e}le and Lallemand, Eliette and Chavatte, Jean-Marc and Crisanti, Andrea and Sinnis, Photini and others},
  journal={PloS One},
  volume={2},
  number={12},
  pages={e1371},
  year={2007},
  publisher={Public Library of Science San Francisco, USA}
}

@article{lopez2017known,
  title={What is known about the immune response induced by Plasmodium vivax malaria vaccine candidates?},
  author={L{\'o}pez, Carolina and Yepes-P{\'e}rez, Yoelis and Hincapi{\'e}-Escobar, Natalia and D{\'\i}az-Ar{\'e}valo, Diana and Patarroyo, Manuel A},
  journal={Frontiers in Immunology},
  volume={8},
  pages={126},
  year={2017},
  publisher={Frontiers Media SA}
}

\end{document}